\documentclass[a4paper,11pt]{article}

\usepackage{jcappub}
\usepackage[utf8]{inputenc}

\usepackage{bm}

\title{Spatially covariant gravity with velocity of the lapse function: the Hamiltonian analysis}

\author{Xian Gao}

\author{and Zhi-Bang Yao}

\affiliation{School of Physics and Astronomy,\\ Sun Yat-sen University, Guangzhou 510275, China}

\emailAdd{gaoxian@mail.sysu.edu.cn}

\emailAdd{yaozhb@mail2.sysu.edu.cn} 

\abstract{
	We investigate a large class of gravity theories that respect spatial covariance, and involve kinetic terms for both the spatial metric and the lapse function. Generally such kind of theories propagate four degrees of freedom, one of which is an unwanted scalar mode. Through a detailed Hamiltonian analysis, we find that the condition requiring the kinetic terms to be degenerate is not sufficient to evade the unwanted scalar mode in general. This is because the primary constraint due to the degeneracy condition does not necessarily induce a secondary constraint, if the mixing terms between temporal and spatial derivatives are present. In this case, the second condition that we dub as the consistency condition must be imposed in order to ensure the existence of the secondary constraint and thus to remove the unwanted mode. We also show how our formalism works through an explicit example, in which the degeneracy condition is not sufficient and thus the consistency condition must be imposed. 
}


\arxivnumber{1806.02811}

\begin{document}
	
	\maketitle


\section{Introduction}

Scalar-tensor theories play an important role in the study of cosmology and gravity theories.
Phenomenologically, they provide the most popular playground for building cosmological models. 
Theoretically, as a modification of Einstein's General Relativity (GR), scalar-tensor theories provide a framework to examine fundamental nature of gravitation.

During the past decade, tremendous progresses have been made in extending the scope of scalar-tensor theories, in particular, by introducing higher derivatives of the scalar field(s) as well as novel couplings between scalar field(s) and the gravity.
The representative achievements along this line are the $k$-essence \cite{ArmendarizPicon:1999rj,Chiba:1999ka} --- the most general theory for covariant scalar field(s) involving first-derivatives in the Lagrangian, and the Horndeski/galileon theory \cite{Horndeski:1974wa,Deffayet:2011gz,Kobayashi:2011nu} --- the most general
covariant scalar-tensor theory involving up to second derivatives both in the Lagrangian as well as in the
equations of motion.
In the recent a couple of years, interests in this direction have been focused on introducing higher order derivatives of the scalar field(s) but with degeneracies  \cite{Gleyzes:2014dya,Gleyzes:2014qga,Langlois:2015cwa,Langlois:2015skt,Crisostomi:2016tcp,Crisostomi:2016czh,Achour:2016rkg,Ezquiaga:2016nqo,Motohashi:2016ftl,deRham:2016wji,BenAchour:2016fzp,Chagoya:2016inc,Crisostomi:2017aim}, which guarantees the healthiness of the theories in the presence of higher derivatives, in particular, the absence of Ostrogradsky ghost \cite{Ostrogradsky:1850fid,Woodard:2015zca}.

An alternative and yet more powerful approach to the scalar-tensor theories is to construct theories that do not respect the full symmetry of GR, i.e., the spacetime diffeomorphism.
In fact, the ``scalar'' is not necessarily a covariant scalar field, but is essentially an effective scalar-type physical degree of freedom.
It is well-known that for theories with gauge redundancies, new degrees of freedom may arise by reducing the gauge symmetries.
Theories along this line can be traced back to the  effective field theory (EFT) of inflation \cite{Creminelli:2006xe,Cheung:2007st} as well as Ho\v{r}ava gravity \cite{Horava:2009uw,Blas:2009qj}. 
The EFT of inflation has been extensively studied and also applied to dark energy \cite{Creminelli:2008wc,Gubitosi:2012hu,Bloomfield:2012ff,Gleyzes:2013ooa,Bloomfield:2013efa,Gleyzes:2014rba,Gleyzes:2015pma,Gleyzes:2015rua}.
These theories are metric theories that respect only spatial symmetry instead of the full spacetime symmetry. 
Therefore, we may refer to such kinds of theories as subsets of ``spatially covariant gravity''.

When fixing the gauge by choosing the scalar field as the time coordinate (i.e., $t = \phi(t,\vec{x})$, which is dubbed as the ``unitary gauge'' in the literature), the Horndeski Lagrangian can be recast in a form similar to that of the EFT of inflation/dark energy \cite{Gleyzes:2013ooa}. 
Interestingly, by deforming the Horndeski Lagrangian in the unitary gauge \cite{Gleyzes:2014dya}, one may get theories beyond the Horndeski domain, of which the healthiness was proven in  \cite{Gleyzes:2014qga,Lin:2014jga,Deffayet:2015qwa}.

From the point of view of spatial symmetries, it is natural to explore such spatially covariant gravity theories as general as we can to see how far they can extend the scope of scalar-tensor theories. 
This possibility was firstly developed in \cite{Gao:2014soa}, where a general framework for spatially covariant gravity was proposed. The Lagrangian is composed of polynomials of the extrinsic curvature $K_{ij}$ --- which encodes the velocity of spatial metric $\dot{h}_{ij}$ --- with coefficients being generally functions of the lapse function $N$, spatial curvature $R_{ij}$ as well as their spatial derivatives.
The covariant scalar-tensor theory corresponding to that in \cite{Gao:2014soa} goes far beyond previously known theories.
Nevertheless, it has been shown that the theory propagates at most 3 physical degrees of freedom, through Hamiltonian analysis in a perturbative manner \cite{Gao:2014fra} and in a non-perturbative manner \cite{Saitou:2016lvb}. 
The framework has also shown novel features in cosmological applications \cite{Kobayashi:2015gga,Fujita:2015ymn,Yajima:2015xva,Akita:2015mho}.

An important ingredient, however, was omitted in the framework in \cite{Gao:2014soa}, that is the velocity  of the lapse function $\dot{N}$.
Although the lapse function $N$ acts as an auxiliary variable in GR, there is no reason to make such a requirement in our construction.
Indeed, if we try to put the spatial metric $h_{ij}$ and the lapse function $N$ in the equal footing, we should include the velocities of both.
The velocity of the spatial metric $\dot{h}_{ij}$ is encoded in the extrinsic curvature $K_{ij} \equiv \frac{1}{2}\pounds_{\bm{n}} h_{ij}$ with $\bm{n}$ the normal vector of the spatial hypersurfaces.
We thus need to include $\dot{N}$ in terms of $\pounds_{\bm{n}}N$ as well.
In fact, as we shall explain in the next section, both $\pounds_{\bm{n}} h_{ij}$ and $\pounds_{\bm{n}}N$ are natural geometric object in our picture.

Another motivation of including $\dot{N}$ in the theory stems from the study of scalar-tensor theories under field transformations, such as the disformal transformation \cite{Bekenstein:1992pj} or mimetic transformation \cite{Chamseddine:2013kea} (see \cite{Sebastiani:2016ras} for a review on mimetic gravity).
Such transformations can be used to generate scalar-tensor theories with higher order-derivatives without the Ostrogradsky ghosts \cite{Bettoni:2013diz,Zumalacarregui:2013pma,Domenech:2015tca,Crisostomi:2016tcp,Achour:2016rkg,Arroja:2015wpa,Takahashi:2017pje}. 
In particular, the transformed theories typically acquire time derivative of the lapse function in the unitary gauge.

In this work, we will extend the framework in \cite{Gao:2014soa} by including the velocity of the lapse function $\dot{N}$ (or precisely $\pounds_{\bm{n}}N$) as one of the basic ingredients.
Generally, if $\dot{N}$ enters the Lagrangian arbitrarily, the theory will propagate 4 physical degrees of freedom.
Our purpose is to build theories with at most 3 degrees of freedom, we thus need to examine under which conditions the unwanted fourth mode can be evaded\footnote{The fourth mode in our formalism is not necessarily an Ostrogradsky ghost mode, as it is not associated with any higher derivative terms. We refer to the fourth modes as ``unwanted'' simply to make connection with the standard scenarios such as inflation or dark energy models with a single clock.}.
Counting degrees of freedom can be well performed in the Hamiltonian constraint analysis.
The current work is thus devoted to this issue.

The paper is organized as following.
In Sec.\ref{sec:action}, we introduce our action by emphasizing the naturalness of $\dot{N}$ as one of the basic ingredients.
In Sec.\ref{sec:pc_Ham}, we derive the primary constraints and define the canonical and total Hamiltonian. 
In Sec.\ref{sec:deg_con},  one degeneracy condition and one consistency condition are derived in order to ensure the healthiness of the theory.
In Sec.\ref{sec:dof}, we derive the secondary constraints and then count the number of physical degrees of freedom by classifying the constraints.
As an explicit example of how our formalism works, in Sec.\ref{sec:exa}, we construct the Lagrangian which corresponds to the most general scalar-tensor theory that is quadratic in second derivative of the scalar field.
Sec.\ref{sec:concl} concludes.

\section{The action} \label{sec:action}

Let us start from the 4-dimensional point of view in order to see what the essential geometric ingredients are in our construction.

The basic picture in our construction is the 4-dimensional spacetime equipped with a foliation of one-parameter family of spacelike hypersurfaces.
The hypersurfaces are specified to be the hypersurfaces with constant values of some scalar field $\phi$.
There are two geometric (independent of local coordinates) quantities in this picture: the 4d metric $g_{\mu\nu}$, and the timelike vector field $n_{\mu}$ normal to the hypersurfaces.
All other ingredients in our construction must be built of $g_{\mu\nu}$ and $n_{\mu}$.
These include the induced metric $h_{\mu\nu} = g_{\mu\nu}+n_{\mu}n_{\nu}$ and the lapse function $N$, which arises from defining the normal vector $n_{\mu} = -N \nabla_{\mu} \phi$.
Derivatives are also decomposed into the ``intrinsic'' one $\mathrm{D}_{\mu}$ and the ``extrinsic'' one $\pounds_{\bm{n}}$, which are the covariant derivative compatible with the induced metric $h_{\mu\nu}$ and the Lie derivative with respect to the normal vector $n^{\mu}\equiv g^{\mu\nu} n_{\mu}$, respectively.
The basic building blocks are thus
	\[
		\phi, N, h_{\mu\nu}, \quad \text{with derivatives in terms of}\quad \mathrm{D}_{\mu}, \pounds_{\bm{n}}.
	\]
At this point, note the shift-vector $N^{\mu}$ that is familiar in the Arnowitt-Deser-Misner formalism should not be included on its own, of which the existence is simply the result of fixing the time direction, and thus requires more information than the geometric structure of the foliation.
In fact, blindly including the shift vector will inevitably introduce more degrees of freedom\footnote{This may however yield other types of interesting theories such as Lorentz-breaking massive gravity \cite{Comelli:2012vz,Comelli:2013paa,Comelli:2013txa,Comelli:2014xga}.}.

The Lie derivative with respect to the normal vector $\pounds_{\bm{n}}$ plays the role of time derivative in our construction.
In this work, we restrict the Lie derivatives $\pounds_{\bm{n}}$ up to the first order, while allow the spatial covariant derivatives $\mathrm{D}_{\mu}$ to be of arbitrarily higher order. 
The action takes the following general form
	\begin{equation}
	S^{\mathrm{(cov)}}:= \int\mathrm{d}^{4}x\sqrt{-g}\,\mathcal{L}^{\mathrm{(cov)}}\left(\phi,N,h_{\mu\nu},F,K_{\mu\nu},\mathrm{D}_{\mu}\right),\label{S_4d}
	\end{equation}
where $F:= \pounds_{\bm{n}}N$ and $K_{\mu\nu} := \frac{1}{2} \pounds_{\bm{n}} h_{\mu\nu}$ is the extrinsic curvature.
Note $\pounds_{\bm{n}}\phi$ does not appear since $\pounds_{\bm{n}} \phi\equiv 1/N$.
In (\ref{S_4d}), the intrinsic Ricci tensor ${}^{3}\!R_{\mu\nu}$ on the hypersurfaces is implicitly included\footnote{Keep in mind that in 3-dimension, the Riemann curvature is not independent, which is determined by Ricci tensor and Ricci scalar.}.
Throughout this work, we consider the hypersurfaces to be spacelike by requiring the scalar field $\phi$ acquires a timelike gradient.
This fact allows us to fix the time coordinate to be the scalar field itself $t= \phi$, which is dubbed as the ``unitary gauge'' in the literature.
In the unitary gauge, the action becomes
	\begin{equation}
	S^{\mathrm{(u.g.)}}=  \int\mathrm{d}t\mathrm{d}^{3}x\,N\sqrt{h}\,\mathcal{L}\left(t,N,h_{ij},F,K_{ij},\nabla_{i}\right),\label{S_ori}
	\end{equation}
with $h\equiv \det h_{ij}$, $\nabla_{i}$ the covariant derivative compatible with the spatial metric $h_{ij}$, and 
	\begin{equation}
		F \equiv \frac{1}{N}\left(\dot{N}-\pounds_{\vec{N}}N\right),\qquad K_{ij}\equiv \frac{1}{2N}\left(\dot{h}_{ij}-\pounds_{\vec{N}}h_{ij}\right), \label{F_Kij}
	\end{equation}
where a dot denotes the time derivative $\partial_{t}$, $\pounds_{\vec{N}}$ denotes the Lie derivative with respect to the shift-vector $N^{i}$.
Again, the spatial Ricci curvature $R_{ij}$ is implicitly included in (\ref{S_ori}).
At this point, it is clear that the framework proposed in \cite{Gao:2014soa,Gao:2014fra} is a special case of (\ref{S_ori}) by turning off the dependence on $F$.

Generally, the Lagrangian in (\ref{S_ori}) may be highly nonlinear in $F$ and $K_{ij}$, which makes the inversion of velocities $\dot{N}$ and $\dot{h}_{ij}$ in terms of the conjugate momenta impossible (at least no compact expressions).
This problem can be solved by introducing Lagrange multipliers $\Lambda$, $\Lambda^{ij}$ as well as the auxiliary fields $A$ and $B_{ij}$ that are spatial tensors, and rewrite (\ref{S_ori}) to be
	\begin{equation}
	\tilde{S}=S+\int\mathrm{d}t\mathrm{d}^{3}x\,N\sqrt{h}\left[\Lambda\left(F-A\right)+\Lambda^{ij}\left(K_{ij}-B_{ij}\right)\right],\label{S_tld_lm}
	\end{equation}
where and in what follows, $S$ denotes $S^{\mathrm{(u.g.)}}$ defined in (\ref{S_ori}) after replacing $F \rightarrow A$ and $K_{ij} \rightarrow B_{ij}$, that is
	\begin{equation}
	S:=\int\mathrm{d}t\mathrm{d}^{3}x\,N\sqrt{h}\,\mathcal{L}\left(t,N,h_{ij},A,B_{ij},\nabla_{i}\right).\label{S_fin}
	\end{equation}
It is clear that (\ref{S_tld_lm}) together with the equations of motion for $\Lambda$ and $\Lambda^{ij}$ reproduces (\ref{S_ori}). Thus the two actions are equivalent, at least at the classical level.

We are also allowed to solve the Lagrange multipliers $\Lambda$ and $\Lambda^{ij}$ by varying (\ref{S_tld_lm}) with respect to $A$ and $B_{ij}$, which yields
	\begin{equation}
	\Lambda=\frac{1}{N\sqrt{h}}\frac{\delta S}{\delta A},\qquad\Lambda^{ij}=\frac{1}{N\sqrt{h}}\frac{\delta S}{\delta B_{ij}}.\label{lm_sol}
	\end{equation}
Then (\ref{S_tld_lm}) becomes
	\begin{equation}
	\tilde{S}=\int\mathrm{d}t\mathrm{d}^{3}x\,N\sqrt{h}\tilde{\mathcal{L}},\label{S_tld_fin}
	\end{equation}
with
	\begin{equation}
	\tilde{\mathcal{L}}:=\mathcal{L}+\frac{1}{N\sqrt{h}}\frac{\delta S}{\delta A}\left(F-A\right)+\frac{1}{N\sqrt{h}}\frac{\delta S}{\delta B_{ij}}\left(K_{ij}-B_{ij}\right).\label{L_tld_fin}
	\end{equation}
(\ref{S_tld_fin}) together with (\ref{L_tld_fin}) will be the starting point of the following analysis.

At this point, note $\tilde{S}$ depends on in total 17 variables
	\begin{equation}
	\left\{\Phi_{I}\right\}:=\left\{ N^{i},A,B_{ij},N,h_{ij}\right\},\label{var}
	\end{equation}
where the indices $I,J,\cdots$ formally denote different kinds of variables as well as their tensorial indices.
The functional dependence of $\tilde{S}$ on the shift-vector $N^i$ is encoded in $F$ and $K_{ij}$ through the Lie derivative $\pounds_{\vec{N}}$. In particular, $S$ in (\ref{S_fin}) has no functional dependence on $N^{i}$.

\section{Primary constraints and the Hamiltonian} \label{sec:pc_Ham}

\subsection{Primary constraints}

The conjugate momenta corresponding to the variables (\ref{var}) are defined as
	\begin{equation}
	\Pi^{I}:=\frac{\delta\tilde{S}}{\delta\dot{\Phi}_{I}} ,\label{mom_def}
	\end{equation}
which are explicitly given by
	\begin{eqnarray}
	\pi_{i} & := & \frac{\delta\tilde{S}}{\delta\dot{N}^{i}}=0,\label{pi_i}\\
	p & := & \frac{\delta\tilde{S}}{\delta\dot{A}}=0,\qquad p^{ij}:=\frac{\delta\tilde{S}}{\delta\dot{B}_{ij}}=0,\label{pPp^ij}\\
	\pi & := & \frac{\delta\tilde{S}}{\delta\dot{N}}=\frac{1}{N}\frac{\delta S}{\delta A},\qquad\pi^{ij} :=\frac{\delta\tilde{S}}{\delta\dot{h}_{ij}} =\frac{1}{2N}\frac{\delta S}{\delta B_{ij}}.\label{pi_pi^ij}
	\end{eqnarray}
According to (\ref{pi_i})-(\ref{pi_pi^ij}), there are in total 17 primary constraints
	\begin{eqnarray}
	\pi_{i} & \approx & 0,\qquad p\approx0,\qquad p^{ij}\approx0,\label{pri_cons_1}\\
	\tilde{\pi} & := & \pi-\frac{1}{N}\frac{\delta S}{\delta A}\approx0,\qquad\tilde{\pi}^{ij}:=\pi^{ij}-\frac{1}{2N}\frac{\delta S}{\delta B_{ij}}\approx 0,\label{pri_cons_2}
	\end{eqnarray}
where and throughout this work ``$\approx$'' represents ``weak equality'' that holds on the subspace in the phase space defined by the primary constraints $\Gamma_{\mathrm{P}}$.
For later convenience, we denote
	\begin{equation}
	\left\{\Pi^{I}\right\} \equiv \left\{ \pi_{i},p,p^{ij},\pi,\pi^{ij}\right\}, \label{mom}
	\end{equation}
for the set of momenta, and
	\begin{equation}
	\left\{\varphi^{I}\right\}:=\left\{ \pi_{i},p,p^{ij},\tilde{\pi},\tilde{\pi}^{ij}\right\},\label{cons_pri}
	\end{equation}
for the set of primary constraints.

\subsection{The Hamiltonian}

The canonical Hamiltonian is obtained by performing a Legendre transformation
	\begin{equation}
	H_{\mathrm{C}}:=\int\mathrm{d}^{3}x\left(\sum_{I}\Pi^{I}\dot{\Phi}_{I}-N\sqrt{h}\tilde{\mathcal{L}}\right), \label{Ham_C_def}
	\end{equation}
where $\tilde{\mathcal{L}}$ is given in (\ref{L_tld_fin}).
Simple manipulation yields
	\begin{eqnarray}
	H_{\mathrm{C}} \approx \left.H_{\mathrm{C}}\right|_{\Gamma_{\mathrm{P}}} & = & \int\mathrm{d}^{3}x\left(\pi\dot{N}+\pi^{ij}\dot{h}_{ij}-N\sqrt{h}\tilde{\mathcal{L}}\right),\nonumber\\
	& = & \int\mathrm{d}^{3}x\left(NC+\pi\pounds_{\vec{N}}\,N+\pi^{ij}\pounds_{\vec{N}}\,h_{ij}\right) , \label{Ham_C_naive}
	\end{eqnarray}
where $\pounds_{\vec{N}}$ is the Lie derivative with respect to the shift vector $N^{i}$, and we define 
	\begin{equation} 
	C:=\pi A+2\pi^{ij}B_{ij}-\sqrt{h}\mathcal{L},\label{C_def}
	\end{equation}
for short with $\mathcal{L}$ given in (\ref{S_fin}).
Since the canonical Hamiltonian is well-defined only on the subspace defined by the primary constraints $\Gamma_{\mathrm{P}}$, one is free to define a new $H_{\mathrm{C}}$ by appending an arbitrary linear combination of primary constraints to $\left.H_{\mathrm{C}}\right|_{\Gamma_{\mathrm{P}}}$ in (\ref{Ham_C_naive}).
For our purpose, we choose
	\begin{equation}
	H_{\mathrm{C}} =  \left.H_{\mathrm{C}}\right|_{\Gamma_{\mathrm{P}}}+\int\mathrm{d}^{3}x\left(p\,\pounds_{\vec{N}}A+p^{ij}\pounds_{\vec{N}}B_{ij}\right), \label{Ham_c_new}
	\end{equation}
of which the reason will become clear soon.
It is then convenient to define the following functional
	\begin{equation}
	X[\vec{N}]:=\int\mathrm{d}^{3}x\sum_{I}\Pi^{I}\pounds_{\vec{N}}\,\Phi_{I},\label{X_Ni_def}
	\end{equation}
where $\vec{N}$ is shift vector, the summation runs over \emph{all} pairs of canonical variables in (\ref{var}) and (\ref{mom_def}).
Using $\pi_{i}\pounds_{\vec{N}}N^i\equiv 0$, (\ref{X_Ni_def}) can be explicitly expanded to be
	\begin{equation}
	X[\vec{N}]=\int\mathrm{d}^{3}x\left(\pi\pounds_{\vec{N}}\,N+\pi^{ij}\pounds_{\vec{N}}\,h_{ij}+p\,\pounds_{\vec{N}}A+p^{ij}\pounds_{\vec{N}}B_{ij}\right). \label{X_Ni_xpl}
	\end{equation}
In terms of $X[\vec{N}]$, $H_{\mathrm{C}}$ in (\ref{Ham_c_new}) can be recast into an elegant form
	\begin{equation}
	H_{\mathrm{C}}=\int\mathrm{d}^{3}x \left(NC\right)+X[\vec{N}],\label{Ham_C_fin}
	\end{equation}
with $C$ given in (\ref{C_def}), where it is clear that the shift vector $N^{i}$ enters the Hamiltonian linearly.
From now on, we will use $H_{\mathrm{C}}$ defined in (\ref{Ham_C_fin}) as our starting point for the Hamiltonian analysis.

Due to the presence of primary constraints, the time-evolution is determined by the so-called total Hamiltonian defined by
	\begin{eqnarray}
	H_{\mathrm{T}} & := & H_{\mathrm{C}}+\int\mathrm{d}^{3}x\left(\lambda_{i}\pi^{i} +v\,p+v_{ij}p^{ij}+\lambda\,\tilde{\pi}+\lambda_{ij}\tilde{\pi}^{ij}\right)\nonumber \\
	& \equiv & H_{\mathrm{C}}+\int\mathrm{d}^{3}y\sum_{I}\lambda_{I}\left(\vec{y}\right)\varphi^{I}\left(\vec{y}\right),\label{Ham_total}
	\end{eqnarray}
where $\{\lambda_{I}\} \equiv \left\{ \lambda_{i},v,v_{ij},\lambda,\lambda_{ij}\right\} $ are undetermined Lagrange multipliers.
In terms of the total Hamiltonian, the time evolution of any phase space function $Q$ is given by
\begin{equation}
\frac{\mathrm{d}Q}{\mathrm{d}t}\approx\frac{\partial Q}{\partial t}+\left[Q,H_{\mathrm{T}}\right], \label{td_pb}
\end{equation}
where $\left[\mathcal{F},\mathcal{G}\right]$ is the Poisson bracket defined by
\begin{equation}
\left[\mathcal{F},\mathcal{G}\right]:=\int\mathrm{d}^{3}z\sum_{I}\left(\frac{\delta\mathcal{F}}{\delta\Phi_{I}\left(\vec{z}\right)}\frac{\delta\mathcal{G}}{\delta\Pi^{I}\left(\vec{z}\right)}-\frac{\delta\mathcal{F}}{\delta\Pi^{I}\left(\vec{z}\right)}\frac{\delta\mathcal{G}}{\delta\Phi_{I}\left(\vec{z}\right)}\right).\label{PB_def}
\end{equation}

For simplicity, from now on we assume that the Lagrangian $\mathcal{L}$ in (\ref{S_fin}) has no explicit time-dependence, i.e., $\partial\mathcal{L}/\partial {t} \equiv 0$. As a result, (\ref{td_pb}) simplifies to be
\begin{equation}
\frac{\mathrm{d}Q}{\mathrm{d}t}\approx \left[Q,H_{\mathrm{T}}\right]. \label{td_pb_s}
\end{equation}
The generalization of the following analysis to a time-dependent Lagrangian is straightforward.

\subsection{More on $X[\vec{N}]$}

Before proceeding, let us discuss some properties of the functional $X[\vec{N}]$.
We may generalize the functional (\ref{X_Ni_def}) to the case with a general spatial vector field $\vec{\xi}$
	\begin{eqnarray}
	X[\vec{\xi}] & := & \int\mathrm{d}^{3}x\sum_{I}\Pi^{I}\pounds_{\vec{\xi}}\,\Phi_{I}\nonumber \\
	& = & \int\mathrm{d}^{3}x\left(\pi_{i}\pounds_{\vec{\xi}}\,N^{i}+\pi\pounds_{\vec{\xi}}\,N+\pi^{ij}\pounds_{\vec{\xi}}\,h_{ij}+p\,\pounds_{\vec{\xi}}A+p^{ij}\pounds_{\vec{\xi}}B_{ij}\right),\label{X_xi_def}
	\end{eqnarray}
where $\vec{\xi}$ may or may not depend on the phase space variables $\{\Phi_{I}, \Pi^{I}\}$. 
For later convenience, note by integration-by-parts, $X[\vec{\xi}]$ defined in (\ref{X_xi_def}) can be recast into
	\begin{equation}
	X[\vec{\xi}]\simeq\int\mathrm{d}^{3}x\,\xi^{i}\mathcal{C}_{i}.\label{X_xi_ibp}
	\end{equation}
with
	\begin{eqnarray}
	\mathcal{C}_{i} & = & \pi\nabla_{i}N-2\sqrt{h}\nabla_{j}\left(\frac{1}{\sqrt{h}}\pi_{i}^{j}\right)+p\nabla_{i}A+p^{kl}\nabla_{i}B_{kl}-2\sqrt{h}\nabla_{j}\left(\frac{p^{jk}}{\sqrt{h}}B_{ik}\right)\nonumber \\
	&  & +\pi_{j}\nabla_{i}N^{j}+\sqrt{h}\nabla_{j}\left(\frac{1}{\sqrt{h}}\pi_{i}N^{j}\right).\label{calCi_def}
	\end{eqnarray}
As a result, the canonical Hamiltonian $H_{\mathrm{C}}$ in (\ref{Ham_C_fin}) can be recast into a more familiar form\footnote{Note the second line in (\ref{calCi_def}) does not contribute to $H_{\mathrm{C}}$ as it will contribute a boundary term. We prefer to used the expression (\ref{calCi_def}) as it is the most convenient form for evaluating the Poisson brackets.}
	\begin{equation}
	H_{\mathrm{C}} \simeq\int\mathrm{d}^{3}x\left(NC+N^{i}\mathcal{C}_{i}\right), \label{Ham_C_ibp}
	\end{equation}
which takes the ``traditional'' form as in General Relativity, where $\mathcal{C}_{i}\rightarrow-2\sqrt{h}\nabla_{j}\left(\frac{1}{\sqrt{h}}\pi_{i}^{j}\right)$ are the momentum constraints.
As we shall see later, $\mathcal{C}_{i}\approx 0$ are nothing but the secondary constraints associated with the primary constraints $\pi_{i}\approx 0$ in our theory.

Comparing with (\ref{Ham_C_ibp}), $H_{\mathrm{C}}$ in the form (\ref{Ham_C_fin}) will be more convenient in the Hamiltonian analysis.
The reason of defining the canonical Hamiltonian as in (\ref{Ham_C_fin}) through the functional $X[\vec{N}]$ is due to the following important property of the more general functional $X[\vec{\xi}]$.
For any functional on the phase space $\mathcal{F} = \mathcal{F}[\Phi_{I},\Pi^{I};q_a]$ that is invariant under the time-independent spatial diffeomorphism, where $\{q_a\}$ stands for variables independent of $\{\Phi_{I},\Pi^{I}\}$, the following equality holds
	\begin{equation}
	\left[\int\mathrm{d}^{3}x\sum_{I}\Pi^{I}\pounds_{\vec{\xi}}\,\Phi_{I},\mathcal{F}\right]=\int\mathrm{d}^{3}x\sum_{I}\Pi^{I}\pounds_{[\vec{\xi},\mathcal{F}]}\Phi_{I} +\int\mathrm{d}^{3}x\sum_{a}\frac{\delta\mathcal{F}}{\delta q_{a}}\pounds_{\vec{\xi}}\,q_{a},\label{PB_LD_xi}
	\end{equation}
or compactly
	\begin{equation}
	\left[X[\vec{\xi}],\mathcal{F}\right]=X\left[[\vec{\xi},\mathcal{F}]\right] +\int\mathrm{d}^{3}x\sum_{a}\frac{\delta\mathcal{F}}{\delta q_{a}}\pounds_{\vec{\xi}}\,q_{a},\label{PB_LD_xi_com}
	\end{equation}
up to a boundary term.
We refer to Appendix \ref{app:PB_LD} for a more general proof.
(\ref{PB_LD_xi}) generalizes similar results in \cite{Gao:2014fra,Mukohyama:2015gia,Saitou:2016lvb}.
As we shall see, introducing $X[\vec{N}]$ in the canonical Hamiltonian (\ref{Ham_C_fin}) significantly simplifies the calculation of Poisson brackets.

Expressions similar to (\ref{calCi_def})  also arise in previous Hamiltonian analysis \cite{Gao:2014fra,Takahashi:2017pje,Lin:2017oow}.
Although $\mathcal{C}_{i}$ in (\ref{calCi_def}) looks complicated, its form will significantly simplify the calculation of Poisson brackets.
In fact, thanks to the property (\ref{PB_LD_xi}) or (\ref{PB_LD_xi_com}), we can make a very strong claim that $\mathcal{C}_{i}$ defined in (\ref{calCi_def}) have vanishing Poisson brackets with any quantity that is weekly vanishing on the phase space. 
Supposing $Q(\vec{x})$ is generally a spatial tensor field or density, where the tensorial indices are suppressed for simplicity. First we can always  build a scalar functional $\mathcal{F}$ out of $Q(\vec{x})$ through certain smoothing function $f(\vec{x})$ (with tensorial indices suppressed as well).
Then using (\ref{X_xi_ibp}) 
with $\vec{\xi}$ some smoothing vector field and (\ref{PB_LD_xi_com}),
	\begin{equation}
	\left[\int\!\mathrm{d}^{3}x\,\xi^{i}(\vec{x})\mathcal{C}_{i}(\vec{x}),\mathcal{F}\right] \simeq \left[X[\vec{\xi}],\mathcal{F}\right]=\int\mathrm{d}^{3}x\,Q\left(\vec{x}\right)\pounds_{\vec{\xi}}\,f\left(\vec{x}\right), \label{PB_calCi_calF}
	\end{equation}
where we have used $[\vec{\xi},\mathcal{F}] \equiv 0$ since $\vec{\xi}$ has nothing to do with the phase space variables.
It immediately follows from (\ref{PB_calCi_calF}) that 
	\begin{equation}
	\left[\mathcal{C}_{i}(\vec{x}),Q(\vec{y})\right] \approx 0\qquad \text{for any}\quad Q\approx 0. \label{PB_calCi_gen}
	\end{equation}
We emphasize that in deriving (\ref{PB_calCi_calF}) and (\ref{PB_calCi_gen}), we have \emph{not} yet used the fact that $\mathcal{C}_{i}$ themselves are constraints.
(\ref{PB_calCi_gen}) with $\mathcal{C}_{i}$ defined in (\ref{calCi_def}) is one of the main results in this work.

\section{Degeneracy and consistency conditions} \label{sec:deg_con}

Constraints must be preserved in time. For the primary constraints $\varphi^{I}$ in (\ref{cons_pri}), we must require that $\mathrm{d}\varphi^{I}/\mathrm{d}t\equiv0$.
According to (\ref{td_pb_s}), it implies
	\begin{equation}
	\int\mathrm{d}^{3}y\sum_{J}\left[\varphi^{I}\left(\vec{x}\right),\varphi^{J}\left(\vec{y}\right)\right]\lambda_{J}\left(\vec{y}\right)+\left[\varphi^{I}\left(\vec{x}\right),H_{\mathrm{C}}\right] \approx 0,\label{cc_pc}
	\end{equation}
which are the consistency conditions for the primary constraints.
(\ref{cc_pc}) may either impose restrictions on the Lagrange multipliers $\{\lambda_{I}\}$'s or reduce to relations among the phase space variables $\{\Phi_{I}, \Pi^{I}\}$.
In the later case, there may be secondary constraints if the relations are independent of the primary constraints.
We thus need to evaluate the Poisson brackets among the primary constraints as well as the Poisson brackets between the primary constraints and the canonical Hamiltonian $H_{\mathrm{C}}$.
In the following we summarize the final results, and refer to Appendix \ref{app:PB_pc} for more details.

For the Poisson brackets among the primary constraints, the following ones are found to be vanishing: 
	\begin{equation}
	\begin{aligned} & \left[\pi_{i}\left(\vec{x}\right),\pi_{j}\left(\vec{y}\right)\right]=0,\\
	& \left[\pi_{i}\left(\vec{x}\right),\tilde{\pi}\left(\vec{y}\right)\right]=0,\qquad\left[\pi_{i}\left(\vec{x}\right),\tilde{\pi}^{kl}\left(\vec{y}\right)\right]=0\\
	& \left[\pi_{i}\left(\vec{x}\right),p\left(\vec{y}\right)\right]=0,\qquad\left[\pi_{i}\left(\vec{x}\right),p^{kl}\left(\vec{y}\right)\right]=0,
	\end{aligned} \label{PB_pi_i_pc}
	\end{equation}
which means that $\pi_{i}$ has vanishing Poisson brackets with all the primary constraints, and
	\begin{eqnarray}
	\left[p\left(\vec{x}\right),p\left(\vec{y}\right)\right] & = & 0,\qquad\left[p\left(\vec{x}\right),p^{ij}\left(\vec{y}\right)\right]=0,\qquad\left[p^{ij}\left(\vec{x}\right),p^{kl}\left(\vec{y}\right)\right]=0.
	\end{eqnarray}
The non-vanishing Poisson brackets are:
	\begin{eqnarray}
	\left[p\left(\vec{x}\right),\tilde{\pi}\left(\vec{y}\right)\right] & = & \frac{1}{N\left(\vec{y}\right)}\frac{\delta^{2}S}{\delta A\left(\vec{x}\right)\delta A\left(\vec{y}\right)},\label{PB_p_pitld}\\
	\left[p\left(\vec{x}\right),\tilde{\pi}^{kl}\left(\vec{y}\right)\right] & = & \frac{1}{2N\left(\vec{y}\right)}\frac{\delta^{2}S}{\delta A\left(\vec{x}\right)\delta B_{kl}\left(\vec{y}\right)},\label{PB_p_pitldkl}\\
	\left[p^{ij}\left(\vec{x}\right),\tilde{\pi}\left(\vec{y}\right)\right] & = & \frac{1}{N\left(\vec{y}\right)}\frac{\delta^{2}S}{\delta B_{ij}\left(\vec{x}\right)\delta A\left(\vec{y}\right)},\label{PB_pij_pitld}\\
	\left[p^{ij}\left(\vec{x}\right),\tilde{\pi}^{kl}\left(\vec{y}\right)\right] & = & \frac{1}{2}\frac{1}{N\left(\vec{y}\right)}\frac{\delta^{2}S}{\delta B_{ij}\left(\vec{x}\right)\delta B_{kl}\left(\vec{y}\right)},\label{PB_pij_pitldkl}
	\end{eqnarray}
and
	\begin{eqnarray}
	\left[\tilde{\pi}\left(\vec{x}\right),\tilde{\pi}\left(\vec{y}\right)\right] & = & \frac{1}{N\left(\vec{y}\right)}\frac{\delta^{2}S}{\delta N\left(\vec{x}\right)\delta A\left(\vec{y}\right)}-\frac{1}{N\left(\vec{x}\right)}\frac{\delta^{2}S}{\delta A\left(\vec{x}\right)\delta N\left(\vec{y}\right)},\\
	\left[\tilde{\pi}\left(\vec{x}\right),\tilde{\pi}^{ij}\left(\vec{y}\right)\right] & = & -\frac{1}{2}\delta^{3}\left(\vec{x}-\vec{y}\right)\frac{1}{N^{2}\left(\vec{y}\right)}\frac{\delta S}{\delta B_{ij}\left(\vec{y}\right)}\nonumber \\
	&  & +\frac{1}{2}\frac{1}{N\left(\vec{y}\right)}\frac{\delta^{2}S}{\delta N\left(\vec{x}\right)\delta B_{ij}\left(\vec{y}\right)}-\frac{1}{N\left(\vec{x}\right)}\frac{\delta^{2}S}{\delta A\left(\vec{x}\right)\delta h_{ij}\left(\vec{y}\right)},\\
	\left[\tilde{\pi}^{ij}\left(\vec{x}\right),\tilde{\pi}^{kl}\left(\vec{y}\right)\right] & = & \frac{1}{2N\left(\vec{y}\right)}\frac{\delta^{2}S}{\delta h_{ij}\left(\vec{x}\right)\delta B_{kl}\left(\vec{y}\right)}-\frac{1}{2N\left(\vec{x}\right)}\frac{\delta^{2}S}{\delta B_{ij}\left(\vec{x}\right)\delta h_{kl}\left(\vec{y}\right)},
	\end{eqnarray}
where $S$ is given in (\ref{S_fin}).

For the Poisson brackets of the primary constraints with $H_{\mathrm{C}}$, we find that
	\begin{equation}
	\left[p\left(\vec{x}\right),H_{\mathrm{C}}\right]\approx0,\qquad\left[p^{ij}\left(\vec{x}\right),H_{\mathrm{C}}\right]\approx 0, \label{PB_p_p^ij_Hc}
	\end{equation}
and the non-vanishing ones are
	\begin{equation}
	\left[\pi_{i}(\vec{x}),H_{\mathrm{C}}\right]=-\mathcal{C}_{i}(\vec{x}), \label{PB_pi_i_Hc}
	\end{equation}
where $\mathcal{C}_i(\vec{x})$ is defined in (\ref{calCi_def}), and
	\begin{eqnarray}
	\left[\tilde{\pi}(\vec{x}),H_{\mathrm{C}}\right] & \approx  & \frac{\delta S}{\delta N(\vec{x})} - \frac{1}{N(\vec{x})}\frac{\delta S}{\delta B_{ij}(\vec{x})}B_{ij}(\vec{x})\nonumber \\
	&  & -\frac{1}{N(\vec{x})}\int\!\mathrm{d}^{3}y\,N(\vec{y})\left(\frac{\delta^{2}S}{\delta A(\vec{x})\delta N(\vec{y})}A(\vec{y})+\frac{\delta^{2}S}{\delta A(\vec{x})\delta h_{ij}(\vec{y})}2B_{ij}(\vec{y})\right),\label{PB_pitld_HC}
	\end{eqnarray}
and
	\begin{eqnarray}
	\left[\tilde{\pi}^{ij}(\vec{x}),H_{\mathrm{C}}\right] & = & \frac{\delta S}{\delta h_{ij}(\vec{x})}+\frac{1}{2}\frac{A(\vec{x})}{N(\vec{x})}\frac{\delta S}{\delta B_{ij}(\vec{x})}\nonumber \\
	&  & -\frac{1}{2}\frac{1}{N(\vec{x})}\int\mathrm{d}^{3}y\,N(\vec{y})\left(\frac{\delta^{2}S}{\delta B_{ij}(\vec{x})\delta N(\vec{y})}A(\vec{y})+\frac{\delta^{2}S}{\delta B_{ij}(\vec{x})\delta h_{kl}(\vec{y})}2B_{kl}(\vec{y})\right).\qquad \label{PB_pitld^ij_HC}
	\end{eqnarray}

Now let us analyse the consistency conditions for the primary constraints (\ref{cc_pc}).
It is convenient to think of (\ref{cc_pc}) in the matrix form and make the following split:
	\begin{equation}
	\varphi^{I}\rightarrow\left(\begin{array}{c}
	\pi_{i}\\
	\varphi^{\alpha}
	\end{array}\right),\qquad\mathrm{and}\qquad\lambda_{I}\rightarrow\left(\begin{array}{c}
	\lambda_{i}\\
	\lambda_{\alpha}
	\end{array}\right),
	\end{equation}
with
	\begin{equation}
	\varphi^{\alpha}\rightarrow\left(\begin{array}{c}
	p\\
	p^{ij}\\
	\tilde{\pi}\\
	\tilde{\pi}^{ij}
	\end{array}\right),\qquad\mathrm{and}\qquad\lambda_{\alpha}\rightarrow\left(\begin{array}{c}
	v\\
	v_{ij}\\
	\lambda\\
	\lambda_{ij}
	\end{array}\right). \label{vphi_lbd_alpha}
	\end{equation}
Accordingly, the Poisson brackets among the primary constraints can be written in the matrix form as
	\begin{equation}
	\left[\varphi^{I}\left(\vec{x}\right),\varphi^{J}\left(\vec{y}\right)\right]\rightarrow\left(\begin{array}{cc}
	\left[\pi_{i}\left(\vec{x}\right),\pi_{k}\left(\vec{y}\right)\right] & \left[\pi_{i}\left(\vec{x}\right),\varphi^{\beta}\left(\vec{y}\right)\right]\\
	\left[\varphi^{\alpha}\left(\vec{x}\right),\pi_{k}\left(\vec{y}\right)\right] & \left[\varphi^{\alpha}\left(\vec{x}\right),\varphi^{\beta}\left(\vec{y}\right)\right]
	\end{array}\right)=\left(\begin{array}{cc}
	\bm{0}_{ik} & \bm{0}_{i}^{\phantom{i}\beta}\\
	\bm{0}_{\phantom{\alpha}k}^{\alpha} & \mathcal{P}^{\alpha\beta}\left(\vec{x},\vec{y}\right)
	\end{array}\right), \label{PB_pc_split}
	\end{equation}
where we have used (\ref{PB_pi_i_pc}) and define
	\begin{equation}
	\mathcal{P}^{\alpha\beta}\left(\vec{x},\vec{y}\right):=\left[\varphi^{\alpha}\left(\vec{x}\right),\varphi^{\beta}\left(\vec{y}\right)\right], \label{calP_def}
	\end{equation}
for later convenience.
In (\ref{PB_pc_split}) and what follows, notation such as $\bm{0}_{ik}$ denotes matrix with null entries. 
Similarly, for the Poisson brackets of the primary constraints with $H_{\mathrm{C}}$ we write
	\begin{equation}
	\left[\varphi^{I}\left(\vec{x}\right),H_{\mathrm{C}}\right]\rightarrow\left(\begin{array}{c}
	\left[\pi_{i}\left(\vec{x}\right),H_{\mathrm{C}}\right]\\
	\left[\varphi^{\alpha}\left(\vec{x}\right),H_{\mathrm{C}}\right]
	\end{array}\right)=\left(\begin{array}{c}
	-\mathcal{C}_{i}\left(\vec{x}\right)\\
	\left[\varphi^{\alpha}\left(\vec{x}\right),H_{\mathrm{C}}\right]
	\end{array}\right),
	\end{equation}
where we used (\ref{PB_pi_i_Hc}) with $\mathcal{C}_{i}$ defined in (\ref{calCi_def}).

With the above split, (\ref{cc_pc}) now takes the form
	\begin{equation}
	\int\mathrm{d}^{3}y\left(\begin{array}{cc}
	\bm{0}_{ik} & \bm{0}_{i}^{\phantom{i}\beta}\\
	\bm{0}_{\phantom{\alpha}k}^{\alpha} & \mathcal{P}^{\alpha\beta}\left(\vec{x},\vec{y}\right)
	\end{array}\right)\left(\begin{array}{c}
	\lambda_{k}\left(\vec{y}\right)\\
	\lambda_{\beta}\left(\vec{y}\right)
	\end{array}\right)+\left(\begin{array}{c}
	-\mathcal{C}_{i}\left(\vec{x}\right)\\
	\left[\varphi^{\alpha}\left(\vec{x}\right),H_{\mathrm{C}}\right]
	\end{array}\right)\approx 0. \label{cc_pc_mtr}
	\end{equation}
The first line in (\ref{cc_pc_mtr}) is simply
	\begin{equation}
		\mathcal{C}_{i} \approx 0,
	\end{equation}
which are the 3 secondary constraints induced by requiring the persistence in time of the primary constraints $\pi_{i}\approx 0$.
The existence of primary constraints $\pi_{i} \approx 0$ together with the associated secondary constraints $\mathcal{C}_{i} \approx 0$ reflects the fact that our theory has spatial covariance.

The second line in (\ref{cc_pc_mtr}) reduces to
	\begin{equation}
	\int\mathrm{d}^{3}y\,\mathcal{P}^{\alpha\beta}\left(\vec{x},\vec{y}\right)\lambda_{\beta}\left(\vec{y}\right)+\left[\varphi^{\alpha}\left(\vec{x}\right),H_{\mathrm{C}}\right]\approx 0. \label{cc_pc_l2}
	\end{equation}
For a general Lagrangian (\ref{S_fin}), the matrix $\mathcal{P}^{\alpha\beta}(\vec{x},\vec{y})$, which consists of Poisson brackets as defined in (\ref{calP_def}), does not degenerate in the sense that 
	\begin{equation}
	\int\mathrm{d}^{3}y\,\mathcal{P}^{\alpha\beta}\left(\vec{x},\vec{y}\right)\mathcal{V}_{\beta}(\vec{y}) = 0. \label{calP_calV}
	\end{equation}
possesses no non-trivial solution for  $\mathcal{V}_{\beta}(\vec{y})$.
In this case, since $\mathcal{P}^{\alpha\beta}(\vec{x},\vec{y})$ is invertible, we can completely determine the Lagrange multipliers $\lambda_{\beta}$ in terms of the phase space variables from (\ref{cc_pc_l2}).
As a result, there is no further secondary constraint to impose, and $\mathcal{C}_{i}\approx 0 $ are the only secondary constraints in our theory.
Let us count the number of physical degrees of freedom in this case. 
To this end, first we note that the secondary constraints $\mathcal{C}_{i} \approx 0$ with $\mathcal{C}_{i}$ defined in (\ref{calCi_def}) have vanishing Poisson brackets with all the primary constraints as well as themselves (see (\ref{PB_calCi_gen}) and discussion there).
According to the terminology of Dirac \cite{Henneaux:1992ig}, we have in total $\#_{\mathrm{1st}} = 6$ first-class constraints $\pi_{i}\approx 0$ and $\mathcal{C}_{i} \approx 0$, and $\#_{\mathrm{2nd}}=14$ second-class constraints $p \approx 0$, $p^{ij}\approx 0$, $\tilde{\pi} \approx 0$ and $\tilde{\pi}^{ij}\approx 0$.
As a result, giving that we have $\#_{\mathrm{var}}=17$ canonical variables, the number of physical degrees of freedom is calculated to be
	\begin{eqnarray}
	\#_{\mathrm{dof}} & = & \frac{1}{2}\left(2\times\#_{\mathrm{var}}-2\times\#_{\mathrm{1st}}-\#_{\mathrm{2nd}}\right)\nonumber \\
	& = & \frac{1}{2}\left(2\times17-2\times6-14\right)\nonumber \\
	& = & 4. \label{num_dof_4}
	\end{eqnarray}
This result implies that besides the two tensor modes that corresponds to the two polarizations of gravitational waves, there are two scalar-type modes getting excited in the theory.
One of these two scalar-type modes is the unwanted mode.

\subsection{Degeneracy condition}

According to the above analysis, in order to get rid of the unwanted mode, we have to impose the degeneracy condition for $\mathcal{P}^{\alpha\beta}$ defined in (\ref{calP_def}), which is the necessary condition to prevent the possible unwanted degree of freedom.
That is, we require that (\ref{calP_calV}) possesses non-trivial solutions $\mathcal{V}_{\beta}(\vec{y}) \neq 0$.
This degeneracy condition will put constraint on the structure of the Lagrangian $\mathcal{L}$ in (\ref{S_fin}).

We make further split of $\varphi^{\alpha}$ and $\lambda_{\alpha}$ in (\ref{vphi_lbd_alpha}) by writing 
	\begin{equation}
	\varphi^{\alpha}\rightarrow\left(\begin{array}{c}
	p^{a}\\
	\tilde{\pi}^{a}
	\end{array}\right),\qquad\lambda_{\alpha}\rightarrow\left(\begin{array}{c}
	v_{a}\\
	\lambda_{a}
	\end{array}\right).
	\end{equation}
with the short-hands
	\begin{equation}
	p^{a}\rightarrow\left(\begin{array}{c}
	p\\
	p^{ij}
	\end{array}\right),\qquad\tilde{\pi}^{a}\rightarrow\left(\begin{array}{c}
	\tilde{\pi}\\
	\tilde{\pi}^{ij}
	\end{array}\right),\qquad v_{a}\rightarrow\left(\begin{array}{c}
	v\\
	v_{ij}
	\end{array}\right),\qquad\lambda_{a}\rightarrow\left(\begin{array}{c}
	\lambda\\
	\lambda_{ij}
	\end{array}\right).
	\end{equation}
Accordingly, $\mathcal{P}^{\alpha\beta}\left(\vec{x},\vec{y}\right)$ defined in (\ref{calP_def}) is decomposed as
	\begin{equation}
	\mathcal{P}^{\alpha\beta}\left(\vec{x},\vec{y}\right) \rightarrow  \left(\begin{array}{cc}
	\left[p^{a}\left(\vec{x}\right),p^{b}\left(\vec{y}\right)\right] & \left[p^{a}\left(\vec{x}\right),\tilde{\pi}^{b}\left(\vec{y}\right)\right]\\
	\left[\tilde{\pi}^{a}\left(\vec{x}\right),p^{b}\left(\vec{y}\right)\right] & \left[\tilde{\pi}^{a}\left(\vec{x}\right),\tilde{\pi}^{b}\left(\vec{y}\right)\right]
	\end{array}\right)
	 \equiv \left(\begin{array}{cc}
	\bm{0}^{ab} & -W^{ba}\left(\vec{y},\vec{x}\right)\\
	W^{ab}\left(\vec{x},\vec{y}\right) & D^{ab}\left(\vec{x},\vec{y}\right)
	\end{array}\right),
	\end{equation}
where we used the fact that (see (\ref{PB_p_p}) and (\ref{PB_p_p^ij}) and the relevant proof there)
	\begin{equation}
		\left[p^{a}\left(\vec{x}\right),p^{b}\left(\vec{y}\right)\right]\approx 0,
	\end{equation}
and defined
	\begin{equation}
	W^{ab}\left(\vec{x},\vec{y}\right):=\left[\tilde{\pi}^{a}\left(\vec{x}\right),p^{b}\left(\vec{y}\right)\right],\qquad D^{ab}\left(\vec{x},\vec{y}\right):=\left[\tilde{\pi}^{a}\left(\vec{x}\right),\tilde{\pi}^{b}\left(\vec{y}\right)\right],\label{WabDab}
	\end{equation}
for short.
Note both $D^{ab}\left(\vec{x},\vec{y}\right)$ and $\mathcal{P}^{\alpha\beta}\left(\vec{x},\vec{y}\right)$ are anti-symmetric in the sense that
	\begin{equation}
	D^{ab}\left(\vec{x},\vec{y}\right)=-D^{ba}\left(\vec{y},\vec{x}\right), \qquad \mathcal{P}^{\alpha\beta}\left(\vec{x},\vec{y}\right) = - \mathcal{P}^{\beta\alpha}\left(\vec{y},\vec{x}\right), \label{DabPab_antisym}
	\end{equation}
which can be verified explicitly.
Generally $W^{ab}\left(\vec{x},\vec{y}\right)$ is not symmetric, although it is always possible to make it symmetric by simple rescale of variables.

In the case of an ordinary matrix $\bm{\mathcal{P}}\equiv\left(\begin{array}{cc}
\bm{0} & -\bm{W}^{\mathrm{T}}\\
\bm{W} & \bm{D}
\end{array}\right)$, the degeneracy condition for $\mathcal{P}^{\alpha\beta}$ is equivalent to the degeneracy condition for $W^{ab}$ since $\det\bm{\mathcal{P}}=\left(\det\bm{W}\right)^{2}$.
This is also true in our case, which can be seen briefly as following.
The degeneracy condition for $\mathcal{P}^{\alpha\beta}$ implies that it possesses at least one null-eigenvector $\mathcal{V}_{\alpha} \neq 0$ satisfying
	\begin{equation}
	\int\mathrm{d}^{3}y\,\mathcal{P}^{\alpha\beta}\left(\vec{x},\vec{y}\right)\mathcal{V}_{\beta}\left(\vec{y}\right) = 0. \label{nullev}
	\end{equation}
By splitting
	\begin{equation}
	\mathcal{V}_{\alpha}\rightarrow\left(\begin{array}{c}
	X_{a}\\
	Y_{a}
	\end{array}\right),
	\end{equation}
(\ref{cc_pc_l2}) yields two equations
	\begin{eqnarray}
	-\int\mathrm{d}^{3}y\,Y_{b}\left(\vec{y}\right)W^{ba}\left(\vec{y},\vec{x}\right) & = & 0,\label{nev_1}\\
	\int\mathrm{d}^{3}y \left[W^{ab}\left(\vec{x},\vec{y}\right)X_{b}\left(\vec{y}\right)+D^{ab}\left(\vec{x},\vec{y}\right)Y_{b}\left(\vec{y}\right)\right] & = & 0,\label{nev_2}
	\end{eqnarray}
where $X_{a}$ and $Y_{a}$ should not be vanishing simultaneously, otherwise $\mathcal{V}_{\alpha} \equiv 0$. 
According to (\ref{nev_1}), there are two cases.
One case is $Y_{a}\equiv0$. In the case, the second line implies $X_{a}\equiv U_{a}\neq0$ must be a null-eigenvector of $W^{ab}\left(\vec{x},\vec{y}\right)$ satisfying
	\begin{equation}
	\int\mathrm{d}^{3}y\,W^{ab}\left(\vec{x},\vec{y}\right)U_{b}\left(\vec{y}\right)\equiv0,\label{nev_U_def}
	\end{equation}
which implies that $W^{ab}$ is degenerate.
The other case is $Y_{a}\neq0$. Then $Y_{a}\equiv V_{a}\neq0$ itself must be a null-eigenvector of $\left(W^{\mathrm{T}}\right)^{ab}\left(\vec{x},\vec{y}\right) \equiv W^{ba}\left(\vec{y},\vec{x}\right)$, satisfying
	\begin{equation}
	\int\mathrm{d}^{3}y\,V_{b}\left(\vec{y}\right)W^{ba}\left(\vec{y},\vec{x}\right)\equiv0,\label{nev_V_def}
	\end{equation}
which also implies that $W_{ab}$ is degenerate.

To summarize, the degeneracy of $W^{ab}$ is a necessary condition
for the degeneracy of $\mathcal{P}^{\alpha\beta}$. As long as $W^{ab}$
is degenerate, and for each null-eigenvector of $W_{ab}$, precisely (\ref{nev_U_def}) or equivalently (\ref{nev_V_def}),
$\mathcal{P}^{\alpha\beta}$ acquires a null-eigenvector corresponding
to
	\begin{equation}
	\mathcal{V}_{\alpha}^{(1)} \equiv \left(\begin{array}{c}
	U_{a}\\
	\bm{0}_{a}
	\end{array}\right),
	\end{equation}
where $U_{a}$ is the eigenvectors of $W^{ab}$ satisfying (\ref{nev_U_def}).
It is possible that $\mathcal{P}^{\alpha\beta}$ acquires a second
null-eigenvector in the form
	\begin{equation}
	\mathcal{V}_{\alpha}^{(2)} \equiv \left(\begin{array}{c}
	X_{a}^{(2)}\\
	V_{a}
	\end{array}\right),
	\end{equation}
with $V_{a}$ the null-eigenvector of $W^{ba}$ satisfying (\ref{nev_V_def}).
However, since $W^{ab}$
is degenerate, in this case it is not guaranteed that (\ref{nev_2}), which now
reads
	\begin{equation}
	\int\mathrm{d}^{3}y\,\left[W^{ab}\left(\vec{x},\vec{y}\right)X_{b}^{(2)}\left(\vec{y}\right)+D^{ab}\left(\vec{x},\vec{y}\right)V_{b}\left(\vec{y}\right)\right]=0,\label{nv_X_eq}
	\end{equation}
always acquires a solution for $X_{a}^{(2)}$. As we shall see later, we
have to impose another consistency condition for $D^{ab}(\vec{x},\vec{y})$
in order to ensure the existence of solution for $X_{a}^{(2)}$ from (\ref{nv_X_eq}).

Let us now focus on the degeneracy condition for $W^{ab}$. 
We make two assumptions:
\begin{itemize}
	\item The nullity of $W^{ab}(\vec{x},\vec{y})$ is $1$.\\
	That is, $W^{ab}(\vec{x},\vec{y})$ has and only has
	one null-eigenvector, satisfying (\ref{nev_U_def}) or equivalently (\ref{nev_V_def}). We emphasize that for each (right) null-eigenvector $U_{a}(\vec{x})$ of $W^{ab}(\vec{x},\vec{y})$ satisfying (\ref{nev_U_def}), there is a corresponding (left) null-eigenvector $V_{a}(\vec{x})$ of $W^{ab}(\vec{x},\vec{y})$ satisfying (\ref{nev_V_def}). In other words, $U_{a}(\vec{x})$ and $V_{a}(\vec{x})$ are not independent.
	\item The degeneracy comes from the $\left\{ \tilde{\pi},p\right\} $-sector.\\
	Precisely, we assume that (using (\ref{PB_pij_pitldkl}))
	\begin{equation}
	\left[\tilde{\pi}^{ij}\left(\vec{x}\right),p^{kl}\left(\vec{y}\right)\right]\propto\frac{\delta^{2}S}{\delta B_{ij}\left(\vec{x}\right)\delta B_{kl}\left(\vec{y}\right)} \label{fd2_BB}
	\end{equation}
	is not degenerate, in the sense that it has no non-trivial null-eigenvector. Recall that $B_{ij}$ is the auxiliary field replacing $K_{ij}$ in the original action (\ref{S_ori}), and thus (\ref{fd2_BB}) implies that the kinetic term for $\dot{h}_{ij}$ is not degenerate, which is a natural assumption and guarantees the theory includes General Relativity as its limit. The degeneracy thus  comes from the manner that $\dot{N}$ enters the original Lagrangian (\ref{S_ori}).
\end{itemize}
In the following we derive the explicit expression of the degeneracy condition for $W_{ab}(\vec{x},\vec{y})$.

According to (\ref{nev_U_def}), it implies two equalities
	\begin{eqnarray}
	\int\mathrm{d}^{3}y\left[\tilde{\pi}\left(\vec{x}\right),p\left(\vec{y}\right)\right]U\left(\vec{y}\right)+\int\mathrm{d}^{3}y\left[\tilde{\pi}\left(\vec{x}\right),p^{kl}\left(\vec{y}\right)\right]U_{kl}\left(\vec{y}\right) & = & 0,\label{WU_1}\\
	\int\mathrm{d}^{3}y\left[\tilde{\pi}^{ij}\left(\vec{x}\right),p\left(\vec{y}\right)\right]U\left(\vec{y}\right)+\int\mathrm{d}^{3}y\left[\tilde{\pi}^{ij}\left(\vec{x}\right),p^{kl}\left(\vec{y}\right)\right]U_{kl}\left(\vec{y}\right) & = & 0.\label{WU_2}
	\end{eqnarray}
Since $\frac{\delta^{2}S}{\delta B_{ij}\left(\vec{x}\right)\delta B_{kl}\left(\vec{y}\right)}$
is not degenerate, it possesses an ``inverse'' $\mathcal{G}_{ij,kl}(\vec{x},\vec{y})$ which is symmetric in the sense
	\begin{equation}
	\ensuremath{\mathcal{G}_{ij,kl}(\vec{x},\vec{y})}=\ensuremath{\mathcal{G}_{kl,ij}(\vec{y},\vec{x})},
	\end{equation}
satisfying
	\begin{equation}
	\int\mathrm{d}^{3}x\,\mathcal{G}_{mn,ij}\left(\vec{z},\vec{x}\right)\frac{\delta^{2}S}{\delta B_{ij}\left(\vec{x}\right)\delta B_{kl}\left(\vec{y}\right)}\equiv\bm{1}_{mn}^{kl}\delta^{3}\left(\vec{z}-\vec{y}\right),\label{calG_fd2_BB}
	\end{equation}
where $\bm{1}_{kl}^{ij}$ is the identity in the linear space of $3\times3$
symmetric matrices satisfying $\bm{1}_{kl}^{ij}h_{ij}\equiv h_{kl}$. 
Equivalently, in terms of $\left[\tilde{\pi}^{ij}\left(\vec{x}\right),p^{kl}\left(\vec{y}\right)\right]$, we have
	\begin{equation}
	-\int\mathrm{d}^{3}x\,\mathcal{G}_{mn,ij}\left(\vec{z},\vec{x}\right)2N\left(\vec{x}\right)\left[\tilde{\pi}^{ij}\left(\vec{x}\right),p^{kl}\left(\vec{y}\right)\right]\equiv\bm{1}_{mn}^{kl}\delta^{3}\left(\vec{z}-\vec{y}\right).\label{calG_pitld_p}
	\end{equation}
Thus we are able to solve $U_{kl}$ in terms of
$U$ from (\ref{WU_2}) to be
	\begin{equation}
	U_{ij}\left(\vec{x}\right)\equiv\int\mathrm{d}^{3}y\,\mathcal{U}_{ij}\left(\vec{x},\vec{y}\right)U\left(\vec{y}\right),\label{Uij_U}
	\end{equation}
with
	\begin{equation}
	\mathcal{U}_{ij}\left(\vec{x},\vec{y}\right):=\int\mathrm{d}^{3}x'\,\mathcal{G}_{ij,kl}\left(\vec{x},\vec{x}'\right)2N\left(\vec{x}'\right)\left[\tilde{\pi}^{kl}\left(\vec{x}'\right),p\left(\vec{y}\right)\right].\label{calU_def}
	\end{equation}
Plugging (\ref{Uij_U}) into (\ref{WU_1}), we get
	\begin{equation}
	-\frac{1}{N\left(\vec{x}\right)}\int\mathrm{d}^{3}y\,\mathcal{D}\left(\vec{x},\vec{y}\right)U\left(\vec{y}\right)=0, \label{dcW_t}
	\end{equation}
with
	\begin{equation}
	\mathcal{D}\left(\vec{x},\vec{y}\right):=-N\left(\vec{x}\right)\left\{ \left[\tilde{\pi}\left(\vec{x}\right),p\left(\vec{y}\right)\right]+\int\mathrm{d}^{3}z\left[\tilde{\pi}\left(\vec{x}\right),p^{ij}\left(\vec{z}\right)\right]\mathcal{U}_{ij}\left(\vec{z},\vec{y}\right)\right\} ,\label{calD_pb}
	\end{equation}
where we deliberately keep the factor $-N(\vec{x})$ for the reason that will become clear soon.
Since (\ref{dcW_t}) should be valid for an arbitrary $U\left(\vec{y}\right) \neq 0$, we must require that
	\begin{equation}
	\mathcal{D}(\vec{x},\vec{y})=0.\label{degen}
	\end{equation}
Using the explicit expressions for the Poisson brackets, we get
	\begin{equation}
	\mathcal{D}(\vec{x},\vec{y})=\frac{\delta^{2}S}{\delta A(\vec{x})\delta A(\vec{y})}-\int\!\mathrm{d}^{3}\vec{x}'\int\!\mathrm{d}^{3}y'\frac{\delta^{2}S}{\delta A(\vec{x})\delta B_{ij}(\vec{x}')}\mathcal{G}_{ij,kl}\left(\vec{x}',\vec{y}'\right)\frac{\delta^{2}S}{\delta B_{kl}(\vec{y}')\delta A(\vec{y})}.\label{calD_def}
	\end{equation}
(\ref{degen}) together with (\ref{calD_def}) is the degeneracy condition for $W^{ab}$, which imposes constraints on the functional dependence of $\mathcal{L}$ in (\ref{S_fin}) on $A$ and $B_{ij}$. In the original Lagrangian (\ref{S_ori}), (\ref{degen}) corresponds to the degeneracy condition for the kinetic terms $\dot{N}$ and $\dot{h}_{ij}$.

For later convenience, note one may alternatively use (\ref{nev_V_def}),
which also yields two equalities
	\begin{eqnarray}
	\int\mathrm{d}^{3}y\,\left[p\left(\vec{x}\right),\tilde{\pi}\left(\vec{y}\right)\right]V\left(\vec{y}\right)+\int\mathrm{d}^{3}y\,\left[p\left(\vec{x}\right),\tilde{\pi}^{kl}\left(\vec{y}\right)\right]V_{kl}\left(\vec{y}\right) & = & 0,\label{WV_1}\\
	\int\mathrm{d}^{3}y\,\left[p^{ij}\left(\vec{x}\right),\tilde{\pi}\left(\vec{y}\right)\right]V\left(\vec{y}\right)+\int\mathrm{d}^{3}y\,\left[p^{ij}\left(\vec{x}\right),\tilde{\pi}^{kl}\left(\vec{y}\right)\right]V_{kl}\left(\vec{y}\right) & = & 0.\label{WV_2}
	\end{eqnarray}
Using the inverse (\ref{calG_pitld_p}), we can solve $V_{kl}$
in terms of $V$ from (\ref{WV_2}), which gives
	\begin{equation}
	V_{ij}\left(\vec{x}\right)\equiv\int\mathrm{d}^{3}y\,\mathcal{V}_{ij}\left(\vec{x},\vec{y}\right)V\left(\vec{y}\right),\label{Vij_V}
	\end{equation}
with
	\begin{equation}
	\mathcal{V}_{ij}\left(\vec{x},\vec{y}\right)\equiv-2N\left(\vec{x}\right)\int\mathrm{d}^{3}y'\,\mathcal{G}_{ij,kl}\left(\vec{x},\vec{y}'\right)\left[p^{kl}\left(\vec{y}'\right),\tilde{\pi}\left(\vec{y}\right)\right].\label{calV_def}
	\end{equation}
Then plugging the solution into (\ref{WV_1}) and using the explicit expressions
for the Poisson brackets, one gets exactly the same degeneracy condition (\ref{degen}) with (\ref{calD_def}).

\subsection{Consistency condition for the existence of the additional secondary constraint} \label{sec:con_sc}

Now let us go back to (\ref{cc_pc_l2}) and discuss under which condition there is an additional secondary constraint.
As we
have discussed after (\ref{nev_2}), as long as $W^{ab}$ is degenerate and of nullity 1,
$\mathcal{P}^{\alpha\beta}$ is degenerate and has one null-eigenvector
	\begin{equation}
	\mathcal{V}_{\alpha}^{(1)}(\vec{x})\equiv\left(\begin{array}{c}
	U_{a}(\vec{x})\\
	\bm{0}_a
	\end{array}\right) \equiv \left(\begin{array}{c}
	U(\vec{x})\\
	U_{ij}(\vec{x})\\
	0\\
	\bm{0}_{ij}
	\end{array}\right)=\int\!\mathrm{d}^{3}y\,\left(\begin{array}{c}
	\delta^{3}(\vec{x}-\vec{y})\\
	\mathcal{U}_{ij}(\vec{x},\vec{y})\\
	0\\
	\bm{0}_{ij}
	\end{array}\right)U(\vec{y}),\label{Pab_nev_1}
	\end{equation}
where we have used the solution for $U_{ij}(\vec{x})$ given in (\ref{Uij_U}) and $\mathcal{U}_{ij}(\vec{x},\vec{y})$ given in (\ref{calU_def}). 
In (\ref{Pab_nev_1}), $U\left(\vec{x}\right)\neq 0$ is
undetermined and can be chosen arbitrarily.
Let us examine whether this null-eigenvector corresponds to a secondary constraint.
To this end, multiplying the left-hand-side of (\ref{cc_pc_l2}) with $\int\mathrm{d}^{3}x\,\mathcal{V}_{\alpha}^{(1)}\left(\vec{x}\right)$ yields
	\begin{eqnarray}
	\mathrm{L.H.S.} & = & \int\mathrm{d}^{3}x\,\mathcal{V}_{\alpha}^{(1)}\left(\vec{x}\right)\left\{ \int\mathrm{d}^{3}y\,\mathcal{P}^{\alpha\beta}\left(\vec{x},\vec{y}\right)\lambda_{\beta}\left(\vec{y}\right)+\left[\varphi^{\alpha}\left(\vec{x}\right),H_{\mathrm{C}}\right]\right\} \nonumber \\
	& = & \int\mathrm{d}^{3}x\,\left(\begin{array}{cc}
	U_{a}\left(\vec{x}\right) & \bm{0}_{a}\end{array}\right)\left(\begin{array}{c}
	\bm{0}^{a}\\
	\left[\tilde{\pi}^{a}\left(\vec{x}\right),H_{\mathrm{C}}\right]
	\end{array}\right)\nonumber \\
	& \equiv & 0,
	\end{eqnarray}
which leads to an identity. In deriving the above, we have used (\ref{PB_p_p^ij_Hc}), i.e., $\left[p^{a}\left(\vec{x}\right),H_{\mathrm{C}}\right]\approx \bm{0}^{a}$.
Thus, if $	\mathcal{V}_{\alpha}^{(1)}$ is the only null-eigenvector of $\mathcal{P}^{\alpha\beta}$, the consistency relations (\ref{cc_pc_l2}) for the primary constraints either are automatically satisfied or simply fix the corresponding Lagrange multipliers.
In particular, this means there is no further secondary constraint. The number of physical degrees of freedom is calculated as that in (\ref{num_dof_4}), but now is $\frac{1}{2}\left(2\times17-2\times7-13\right)=3.5$.
This implies the phase space in each spacetime point is of odd dimension, which is not consistent \cite{Henneaux:2009zb,Li:2009bg}.
Thus the degeneracy condition we derived above is merely a necessary condition for the healthiness of the theory.

Fortunately, it is possible that $\mathcal{P}^{\alpha\beta}$ acquires
a second null-eigenvector in the form
	\begin{equation}
	\mathcal{V}_{\alpha}^{(2)}\equiv\left(\begin{array}{c}
	X_{a}^{(2)}\\
	V_{a}
	\end{array}\right),\label{Pab_nev_2}
	\end{equation}
with $V_{a} \equiv (V,V_{ij})^{\mathrm{T}}$, where $V_{ij}(\vec{x})$ is given in terms of $V(\vec{x})$ as (\ref{Vij_V}) with $V\left(\vec{x}\right)\neq0$
undetermined. 
The crucial point is, in order to have this null-eigenvector
$\mathcal{V}_{\alpha}^{(2)}$ to exist, we have to impose certain conditions for
$D^{ab}\left(\vec{x},\vec{y}\right)$ defined in (\ref{WabDab}).
As we have discussed after (\ref{nv_X_eq}), this condition is simply
the consistency condition for (\ref{nv_X_eq}). That is, (\ref{nv_X_eq})
indeed acquires a solution for $X_{a}^{(2)}$.
In the following, we derive this condition by solving $\mathcal{V}_{\alpha}^{(2)}$ explicitly.

Under the splitting
	\begin{equation}
	X_{a}^{(2)}\rightarrow\left(\begin{array}{c}
	X^{(2)}\\
	X_{ij}^{(2)}
	\end{array}\right),
	\end{equation}
(\ref{nv_X_eq}) yields two equations,
	\begin{eqnarray}
	0 & = & \int\mathrm{d}^{3}y\,\Big\{\left[\tilde{\pi}\left(\vec{x}\right),p\left(\vec{y}\right)\right]X^{(2)}\left(\vec{y}\right)+\left[\tilde{\pi}\left(\vec{x}\right),p^{kl}\left(\vec{y}\right)\right]X_{kl}^{(2)}\left(\vec{y}\right)\nonumber \\
	&  & \quad+\left[\tilde{\pi}\left(\vec{x}\right),\tilde{\pi}\left(\vec{y}\right)\right]V\left(\vec{y}\right)+\left[\tilde{\pi}\left(\vec{x}\right),\tilde{\pi}^{kl}\left(\vec{y}\right)\right]V_{kl}\left(\vec{y}\right)\Big\},\label{X2_eq1}\\
	0 & = & \int\mathrm{d}^{3}y\,\Big\{\left[\tilde{\pi}^{ij}\left(\vec{x}\right),p\left(\vec{y}\right)\right]X^{(2)}\left(\vec{y}\right)+\left[\tilde{\pi}^{ij}\left(\vec{x}\right),p^{kl}\left(\vec{y}\right)\right]X_{kl}^{(2)}\left(\vec{y}\right)\nonumber \\
	&  & \quad+\left[\tilde{\pi}^{ij}\left(\vec{x}\right),\tilde{\pi}\left(\vec{y}\right)\right]V\left(\vec{y}\right)+\left[\tilde{\pi}^{ij}\left(\vec{x}\right),\tilde{\pi}^{kl}\left(\vec{y}\right)\right]V_{kl}\left(\vec{y}\right)\Big\}.\label{X2_eq2}
	\end{eqnarray}
Multiplying (\ref{X2_eq2}) by $-\int\mathrm{d}^{3}x\,\mathcal{G}_{mn,ij}\left(\vec{z},\vec{x}\right)2N\left(\vec{x}\right)$ and using (\ref{calG_pitld_p}) and (\ref{Vij_V})
yield the solution for $X_{kl}^{(2)}$ in terms of $X^{(2)}$ and $V$:
	\begin{equation}
	X_{ij}^{(2)}\left(\vec{x}\right)=\int\mathrm{d}^{3}y\,\mathcal{U}_{ij}\left(\vec{x},\vec{y}\right)X^{(2)}\left(\vec{y}\right)+\int\mathrm{d}^{3}y\,\mathcal{X}_{ij}\left(\vec{x},\vec{y}\right)V\left(\vec{y}\right),\label{X2ij_gen}
	\end{equation}
where $\mathcal{U}_{ij}(\vec{x},\vec{y})$ is defined in (\ref{calU_def}), and $\mathcal{X}_{ij}\left(\vec{x},\vec{y}\right)$ is defined to be
	\begin{eqnarray}
	\mathcal{X}_{ij}\left(\vec{x},\vec{y}\right) & := & \int\mathrm{d}^{3}x'\,\mathcal{G}_{ij,kl}\left(\vec{x},\vec{x}'\right)2N\left(\vec{x}'\right)\nonumber \\
	&  & \quad\times\left\{ \left[\tilde{\pi}^{kl}\left(\vec{x}'\right),\tilde{\pi}\left(\vec{y}\right)\right]+\int\mathrm{d}^{3}y'\,\left[\tilde{\pi}^{kl}\left(\vec{x}'\right),\tilde{\pi}^{mn}\left(\vec{y}'\right)\right]\mathcal{V}_{mn}\left(\vec{y}',\vec{y}\right)\right\},\label{calX_def}
	\end{eqnarray}
with $\mathcal{V}_{ij}(\vec{x},\vec{y})$ given in (\ref{calV_def}). 
With the above solution for $X_{ij}^{(2)}$, we have
	\begin{equation}
	\mathcal{V}_{\alpha}^{(2)}\left(\vec{x}\right)\equiv\left(\begin{array}{c}
	X^{(2)}(\vec{x})\\
	X_{ij}^{(2)}(\vec{x})\\
	V(\vec{x})\\
	V_{ij}(\vec{x})
	\end{array}\right)=\int\!\mathrm{d}^{3}y\left[\left(\begin{array}{c}
	\delta^{3}(\vec{x}-\vec{y})\\
	\mathcal{U}_{ij}(\vec{x},\vec{y})\\
	0\\
	\bm{0}_{ij}
	\end{array}\right)X^{(2)}(\vec{y})+\left(\begin{array}{c}
	0\\
	\mathcal{X}_{ij}(\vec{x},\vec{y})\\
	\delta^{3}(\vec{x}-\vec{y})\\
	\mathcal{V}_{ij}(\vec{x},\vec{y})
	\end{array}\right)V(\vec{y})\right],\label{Pab_nev_2_xpl}
	\end{equation}
which depends on two arbitrary functions $X^{(2)}(\vec{x})$ and $V(\vec{x})$.
At this point, we can make a simplification by observing that the $X^{(2)}$ contribution to $\mathcal{V}_{\alpha}^{(2)}$ in (\ref{Pab_nev_2_xpl}) is exactly of the same form as $\mathcal{V}_{\alpha}^{(1)}$ (see the last equality of (\ref{Pab_nev_1})) and thus is not independent. 
Thus we are free to set $X^{(2)} = 0$, and from now on we choose $X^{(2)} = 0$ in the solutions (\ref{X2ij_gen}) and (\ref{Pab_nev_2_xpl}).

Then there comes the crucial point. We have to make sure that the above solution for $X_{ij}^{(2)}$ is consistent with (\ref{X2_eq1}). Plugging (\ref{X2ij_gen})
into (\ref{X2_eq1}) yields\footnote{In fact if one keeps $X^{(2)} \neq 0$ in the solution (\ref{X2ij_gen}), one will find exactly (\ref{X2_eq1_calF}) as well, since the contribution of $X^{(2)}$ is proportional to the degeneracy condition for $W^{ab}$ in (\ref{degen}) and thus is identically vanishing.}
	\begin{equation}
	\int\mathrm{d}^{3}y\,\mathcal{F}\left(\vec{x},\vec{y}\right)V\left(\vec{y}\right)=0,\label{X2_eq1_calF}
	\end{equation}
where we define
	\begin{eqnarray}
	\mathcal{F}\left(\vec{x},\vec{y}\right) & := & \left[\tilde{\pi}\left(\vec{x}\right),\tilde{\pi}\left(\vec{y}\right)\right]\nonumber \\
	&  & +\int\!\mathrm{d}^{3}z\left\{ \left[\tilde{\pi}\left(\vec{x}\right),\tilde{\pi}^{ij}\left(\vec{z}\right)\right]\mathcal{V}_{ij}\left(\vec{z},\vec{y}\right)-\left[\tilde{\pi}\left(\vec{y}\right),\tilde{\pi}^{ij}\left(\vec{z}\right)\right]\mathcal{V}_{ij}\left(\vec{z},\vec{x}\right)\right\} \nonumber \\
	&  & +\int\!\mathrm{d}^{3}x'\int\!\mathrm{d}^{3}y'\,\mathcal{V}_{ij}\left(\vec{x}',\vec{x}\right)\left[\tilde{\pi}^{ij}\left(\vec{x}'\right),\tilde{\pi}^{kl}\left(\vec{y}'\right)\right]\mathcal{V}_{kl}\left(\vec{y}',\vec{y}\right),\label{calF_def}
	\end{eqnarray}
with $\mathcal{V}_{ij}$ given in (\ref{calV_def}).
Note from (\ref{calF_def}) it is clear that $\mathcal{F}\left(\vec{x},\vec{y}\right)$ is antisymmetric in the sense that
	\begin{equation}
	\mathcal{F}\left(\vec{y},\vec{x}\right)=-\mathcal{F}\left(\vec{x},\vec{y}\right),\label{calF_as}
	\end{equation}
which can be checked easily. Since (\ref{X2_eq1_calF}) must be valid for an arbitrary $V\left(\vec{x}\right) \neq 0$,
we must require that
	\begin{equation}
	\mathcal{F}\left(\vec{x},\vec{y}\right)=0, \label{calF_cond}
	\end{equation}
which is the consistency condition in order to ensure the existence of an additional secondary constraint. 

From now on, we assume (\ref{calF_cond}) is satisfied, and thus $\mathcal{P}^{\alpha\beta}$
possesses two linearly independent null-eigenvectors  $\mathcal{V}^{(1)}_{\alpha}$ in (\ref{Pab_nev_1})
and $\mathcal{V}^{(2)}_{\alpha}$ in (\ref{Pab_nev_2_xpl}) with $X^{(2)} \equiv 0$.

\section{The physical degrees of freedom} \label{sec:dof}

\subsection{The additional secondary constraint}

Before proceeding with revealing the additional secondary constraint, let us consider two  linear combinations of the primary constraints through null-eigenvectors $\mathcal{V}^{(1)}_{\alpha}$ and $\mathcal{V}^{(2)}_{\alpha}$.
The idea is that, as in linear algebra for ordinary matrix, the matrix of Poisson brackets among the primary constraints (\ref{PB_pc_split}) can be reduced by introducing new combinations of the primary constraints through the null-eigenvectors.

For the first null-eigenvector $\mathcal{V}^{(1)}_{\alpha}$, using (\ref{Pab_nev_1}),
	\begin{eqnarray}
	\int\mathrm{d}^{3}x\mathcal{V}_{\alpha}^{(1)}\varphi^{\alpha}\left(\vec{x}\right) & = & \int\mathrm{d}^{3}x\left[U\left(\vec{x}\right)p\left(\vec{x}\right)+U_{ij}\left(\vec{x}\right)p^{ij}\left(\vec{x}\right)\right]\nonumber \\
	& \equiv & \int\mathrm{d}^{3}x\,\bar{p}\left(\vec{x}\right) U\left(\vec{x}\right),\label{pbar_U_nev1}
	\end{eqnarray}
where we have used (\ref{Uij_U}) and define
	\begin{equation}
	\bar{p}\left(\vec{x}\right):=p\left(\vec{x}\right)+\int\mathrm{d}^{3}y\,p^{ij}\left(\vec{y}\right)\mathcal{U}_{ij}\left(\vec{y},\vec{x}\right),\label{pbar_def}
	\end{equation}
with $\mathcal{U}_{ij}$ given in (\ref{calU_def}).
Keep in mind that in (\ref{pbar_U_nev1}), $U\left(\vec{x}\right)\neq 0$ is some undetermined spatial scalar field and can be chosen arbitrarily, which has nothing to
do with phase space variables and thus behaves as a smoothing function.

Similarly, for the second null-eigenvector $\mathcal{V}^{(2)}_{\alpha}$, using (\ref{Pab_nev_2_xpl}) (with $X^{(2)} = 0$),
	\begin{eqnarray}
	\int\mathrm{d}^{3}x\,\mathcal{V}_{\alpha}^{(2)}\left(\vec{x}\right)\varphi^{\alpha}\left(\vec{x}\right) & = & \int\mathrm{d}^{3}x\left[X_{ij}^{(2)}\left(\vec{x}\right)p^{ij}\left(\vec{x}\right)+V\left(\vec{x}\right)\tilde{\pi}\left(\vec{x}\right)+V_{ij}\left(\vec{x}\right)\tilde{\pi}^{ij}\left(\vec{x}\right)\right]\nonumber \\
	& \equiv & \int\mathrm{d}^{3}x\,\bar{\pi}\left(\vec{x}\right)V\left(\vec{x}\right), \label{pibar_V_nev2}
	\end{eqnarray}
where we have used (\ref{X2ij_gen}) and (\ref{Vij_V}), and  define
	\begin{equation}
	\bar{\pi}\left(\vec{x}\right):=\tilde{\pi}\left(\vec{x}\right)+\int\mathrm{d}^{3}y\,p^{ij}\left(\vec{y}\right)\mathcal{X}_{ij}\left(\vec{y},\vec{x}\right)+\int\mathrm{d}^{3}y\,\tilde{\pi}^{ij}\left(\vec{y}\right)\mathcal{V}_{ij}\left(\vec{y},\vec{x}\right),\label{pibar_def}
	\end{equation}
with $\mathcal{X}_{ij}$ and $\mathcal{V}_{ij}$ given in (\ref{calX_def}) and (\ref{calV_def}), respectively.

Since $\bar{p}$ is the linear combination of $p$ and $p^{ij}$, $\bar{\pi}$ is the linear combination of $\tilde{\pi}$, $p^{ij}$ and $\tilde{\pi}^{ij}$, we may use
	\begin{equation}
		\{ \pi_{i}, \bar{p}, p^{ij}, \bar{\pi}, \tilde{\pi}^{ij} \},
	\end{equation}
as our new complete set of primary constraints.

We are now ready to reveal the additional secondary constraint hidden in (\ref{cc_pc_l2}).
As we have discussed in Sec.\ref{sec:con_sc}, the null-eigenvector $\mathcal{V}^{(1)}_{\alpha}$ has no corresponding secondary constraint.
On the other hand, multiplying the left-hand-side of (\ref{cc_pc_l2}) by  $\int\!\mathrm{d}^{3}x\,\mathcal{V}_{\alpha}^{(2)}(\vec{x})$ yields
	\begin{eqnarray}
	\mathrm{L.H.S.} & = & \int\mathrm{d}^{3}x\,\mathcal{V}_{\alpha}^{(2)}\left(\vec{x}\right)\left\{ \int\mathrm{d}^{3}y\,\mathcal{P}^{\alpha\beta}\left(\vec{x},\vec{y}\right)\lambda_{\beta}\left(\vec{y}\right)+\left[\varphi^{\alpha}\left(\vec{x}\right),H_{\mathrm{C}}\right]\right\} \nonumber \\
	& = & \int\mathrm{d}^{3}x\,\mathcal{V}_{\alpha}^{(2)}\left(\vec{x}\right)\left[\varphi^{\alpha}\left(\vec{x}\right),H_{\mathrm{C}}\right],
	\end{eqnarray}
which does not vanish identically, and thus corresponds to a secondary
constraint. 
In fact, the above can be further simplified to be
	\begin{equation}
	\mathrm{L.H.S.}\approx\left[\int\mathrm{d}^{3}x\,\mathcal{V}_{\alpha}^{(2)}\left(\vec{x}\right)\varphi^{\alpha}\left(\vec{x}\right),H_{\mathrm{C}}\right]\equiv\left[\int\mathrm{d}^{3}x\,\bar{\pi}\left(\vec{x}\right)V\left(\vec{x}\right),H_{\mathrm{C}}\right]\approx\int\mathrm{d}^{3}x\,\mathcal{C}\left(\vec{x}\right)V\left(\vec{x}\right),\label{nev2_cc}
	\end{equation}
where $\bar{\pi}(\vec{x})$ is defined in (\ref{pibar_V_nev2}) and we define
	\begin{equation}
	\mathcal{C}\left(\vec{x}\right)\equiv\left[\bar{\pi}\left(\vec{x}\right),H_{\mathrm{C}}\right].\label{calC_def}
	\end{equation}
Using (\ref{pibar_def}), we get
	\begin{equation}
	\mathcal{C}\left(\vec{x}\right)\approx\left[\tilde{\pi}\left(\vec{x}\right),H_{\mathrm{C}}\right]+\int\mathrm{d}^{3}y\left[\tilde{\pi}^{ij}\left(\vec{y}\right),H_{\mathrm{C}}\right]\mathcal{V}_{ij}\left(\vec{y},\vec{x}\right),\label{calC_xpl}
	\end{equation}
Since (\ref{nev2_cc}) must be vanishing for arbitrary $V\left(\vec{x}\right) \neq 0$,
we must require that
	\begin{equation}
	\mathcal{C}\left(\vec{x}\right)\approx 0,
	\end{equation}
which is the additional secondary constraint arising from the persistence in time of $\bar{\pi}(\vec{x}) \approx 0$.

At this point, note from (\ref{PB_calCi_gen}) the consistency condition for the secondary constraint $\mathcal{C}_{i} \approx 0$ is automatically satisfied.
While generally $\left[\mathcal{C}\left(\vec{x}\right),\varphi^{\beta}\left(\vec{y}\right)\right] \neq 0$, which implies the consistency condition for $\mathcal{C}$ simply fixes Lagrange multipliers $\lambda_{\alpha}$ and yields no further secondary constraint.

\subsection{Classification of constraints and the physical degrees of freedom}

To summarize, supposing that our theory is built such that both the degeneracy and consistency conditions (\ref{degen}) and (\ref{calF_cond}) are satisfied, we thus have 17 primary constraints $\{\pi_{i}, \bar{p}, p^{ij}, \bar{\pi}, \tilde{\pi}^{ij} \}$ and 4 secondary constraints $\{\mathcal{C}_{i},\mathcal{C}\}$.
The number of physical degrees of freedom is subject to the classification of all the constraints, which we will discuss below.

It immediately follows that the primary constraints $\pi_{i} \approx 0$ and the secondary constraints $\mathcal{C}_{i} \approx 0$ have vanishing Poisson brackets with all the constraints (see (\ref{PB_calCi_gen})) and thus are first-class, according to Dirac's terminology.
Mathematically, this is because the shift-vector $N^{i}$ enters the Hamiltonian linearly in terms of the Lie derivatives $\pounds_{\vec{N}}$.
Physically, this is due to the fact that our theory has spatial covariance.

The crucial point is, as long as the degeneracy condition (\ref{degen}) and the consistency condition (\ref{calF_cond}) are satisfied, there is  one additional first-class constraint, which is encoded in the linear combination  of the rest constraints.
Now let us reveal this additional first-class constraint by considering the Poisson brackets of the new primary constraint $\bar{p}(\vec{x})$ defined in (\ref{pbar_def}).
It is more convenient to consider the functional in (\ref{pbar_U_nev1}), which is equivalent to
	\begin{equation}
	\bar{p}[U]:=\int\mathrm{d}^{3}x\,\bar{p}\left(\vec{x}\right)U\left(\vec{x}\right)\equiv\int\mathrm{d}^{3}x\,p^{a}\left(\vec{x}\right) U_{a}\left(\vec{x}\right), \label{pbar_U_fn}
	\end{equation}
where $U_{a}$ is the null-eigenvector of $W^{ab}$ defined as in (\ref{nev_U_def}). 
We emphasize that $U\left(\vec{x}\right)\neq 0$ is a spatial scalar field undetermined and can be chosen arbitrarily. In particular, $U\left(\vec{x}\right)$ has nothing to do with phase space variables and thus behaves as a smoothing function. 

It is easy to show that
	\begin{equation}
	\left[p^{a}\left(\vec{x}\right),\bar{p}[U]\right]\approx\int\mathrm{d}^{3}y\,\left[p^{a}\left(\vec{x}\right),p^{b}\left(\vec{y}\right)\right]U_{b}\left(\vec{y}\right)=0, \label{PB_p_Pbar}
	\end{equation}
and
	\begin{equation}
	\left[\tilde{\pi}^{a}\left(\vec{x}\right),\bar{p}[U]\right]\approx\int\mathrm{d}^{3}\vec{y}\,\left[\tilde{\pi}^{a}\left(\vec{x}\right),p^{b}\left(\vec{y}\right)\right]U_{b}\left(\vec{y}\right)=0, \label{PB_pitld_Pbar}
	\end{equation}
where in (\ref{PB_p_Pbar}) we used $\left[p^{a}(\vec{x}),p^{b}(\vec{y})\right]\equiv0$, and in (\ref{PB_pitld_Pbar}) we used the fact that $U_{b}(\vec{x})$ is the null-eigenvector of $W^{ab}(\vec{x},\vec{y})\equiv\left[\tilde{\pi}^{a}(\vec{x}),p^{b}(\vec{y})\right]$.
It thus follows from (\ref{PB_p_Pbar}) and (\ref{PB_pitld_Pbar}) that
	\begin{equation}
		\left[p^{a}\left(\vec{x}\right),\bar{p}\left(\vec{y}\right)\right]\approx 0,\qquad\left[\tilde{\pi}^{a}\left(\vec{x}\right),\bar{p}\left(\vec{y}\right)\right]\approx 0. \label{PB_pbar_pa_pitlda}
	\end{equation}
According to (\ref{PB_pbar_pa_pitlda}) and the definition (\ref{pbar_def}), $\bar{p}(\vec{x})$ also has vanishing Poisson brackets with itself and with $\bar{\pi}(\vec{x})$ defined in (\ref{pibar_def}):
	\begin{equation}
		\left[\bar{p}\left(\vec{x}\right),\bar{p}\left(\vec{y}\right)\right]\approx 0,\qquad \left[\bar{\pi}\left(\vec{x}\right),\bar{p}\left(\vec{y}\right)\right]\approx 0,.
	\end{equation}
The last Poisson bracket we need to check is $\left[\mathcal{C}\left(\vec{x}\right),\bar{p}\left(\vec{y}\right)\right]$.
After some non-trivial manipulations (see Appendix \ref{sec:PB_calC_pbar} for details), we find
	\begin{equation}
	\left[\mathcal{C}\left(\vec{x}\right),\bar{p}\left(\vec{y}\right)\right]\approx 0.
	\end{equation}
To summarize, $\bar{p}(\vec{x})$ defined in (\ref{pbar_def}) has vanishing Poisson brackets with all the constraints and thus is first class.

Generally 
\begin{equation}
\left[\mathcal{C}(\vec{x}),p^{kl}(\vec{y})\right]\neq0,\qquad\left[\mathcal{C}(\vec{x}),\tilde{\pi}^{kl}(\vec{y})\right]\neq0.
\end{equation}
Nevertheless, we can always introduce a further combination
\begin{equation}
\bar{\mathcal{C}}(\vec{x})\equiv\mathcal{C}(\vec{x})+\int\!\mathrm{d}^{3}z\left\{ \mathcal{S}_{ij}\left(\vec{x},\vec{z}\right)p^{ij}(\vec{z})+\mathcal{T}_{ij}\left(\vec{x},\vec{z}\right)\tilde{\pi}^{ij}(\vec{z})\right\} \label{calCbar_def}
\end{equation}
such that
\begin{equation}
\left[\bar{\mathcal{C}}(\vec{x}),p^{kl}(\vec{y})\right]\approx0,\qquad\left[\bar{\mathcal{C}}(\vec{x}),\tilde{\pi}^{kl}(\vec{y})\right]\approx0.
\end{equation}
In fact, from (\ref{calCbar_def}) we have
\begin{equation}
\left[\bar{\mathcal{C}}(\vec{x}),p^{kl}(\vec{y})\right]\approx\left[\mathcal{C}(\vec{x}),p^{kl}(\vec{y})\right]+\int\mathrm{d}^{3}z\,\mathcal{T}_{ij}\left(\vec{x},\vec{z}\right)\left[\tilde{\pi}^{ij}(\vec{z}),p^{kl}(\vec{y})\right],
\end{equation}
and
\begin{eqnarray}
\left[\bar{\mathcal{C}}(\vec{x}),\tilde{\pi}^{kl}(\vec{y})\right] & \approx & \left[\mathcal{C}(\vec{x}),\tilde{\pi}^{kl}(\vec{y})\right]+\int\mathrm{d}^{3}z\,\mathcal{S}_{ij}\left(\vec{x},\vec{z}\right)\left[p^{ij}(\vec{z}),\tilde{\pi}^{kl}(\vec{y})\right]\nonumber \\
&  & +\int\mathrm{d}^{3}z\,\mathcal{T}_{ij}\left(\vec{x},\vec{z}\right)\left[\tilde{\pi}^{ij}(\vec{z}),\tilde{\pi}^{kl}(\vec{y})\right],
\end{eqnarray}
which always acquire solutions for $\mathcal{S}_{ij}$ and $\mathcal{T}_{ij}$.

Finally, the Poisson brackets among all the constraints are summarized
in the following table:
	\begin{center}
		\begin{tabular}{c|ccccccc}
			\hline 
			$\left[\cdot,\cdot\right]$ & $\pi_{k}(\vec{y})$ & $\bar{p}(\vec{y})$ & $p^{kl}(\vec{y})$ & $\bar{\pi}(\vec{y})$ & $\tilde{\pi}^{kl}(\vec{y})$ & $\mathcal{C}_{k}(\vec{y})$ & $\bar{\mathcal{C}}(\vec{y})$\tabularnewline
			\hline 
			$\pi_{i}(\vec{x})$ & $0$ & $0$ & $0$ & $0$ & $0$ & $0$ & $0$\tabularnewline
			$\bar{p}(\vec{x})$ & $0$ & $0$ & $0$ & $0$ & $0$ & $0$ & $0$\tabularnewline
			$p^{ij}(\vec{x})$ & $0$ & $0$ & $0$ & $0$ & $[p^{ij}(\vec{x}),\tilde{\pi}^{kl}(\vec{y})]$ & $0$ & $0$\tabularnewline
			$\bar{\pi}(\vec{x})$ & $0$ & $0$ & $0$ & $0$ & $0$ & $0$ & $[\bar{\pi}(\vec{x}),\bar{\mathcal{C}}(\vec{y})]$\tabularnewline
			$\tilde{\pi}^{ij}(\vec{x})$ & $0$ & $0$ & $[\tilde{\pi}^{ij}(\vec{x}),p^{kl}(\vec{y})]$ & $0$ & $[\tilde{\pi}^{ij}(\vec{x}),\tilde{\pi}^{kl}(\vec{y})]$ & $0$ & $0$\tabularnewline
			$\mathcal{C}_{i}(\vec{x})$ & $0$ & $0$ & $0$ & $0$ & $0$ & $0$ & $0$\tabularnewline
			$\bar{\mathcal{C}}(\vec{x})$ & $0$ & $0$ & $0$ & $[\bar{\mathcal{C}}(\vec{x}),\bar{\pi}(\vec{y})]$ & $0$ & $0$ & $[\bar{\mathcal{C}}(\vec{x}),\bar{\mathcal{C}}(\vec{y})]$\tabularnewline
			\hline 
		\end{tabular}
		\par\end{center}
Generally,
	\begin{equation}
	\left[\bar{\mathcal{C}}(\vec{x}),\bar{\pi}(\vec{y})\right]\neq0,\qquad\left[\bar{\mathcal{C}}(\vec{x}),\bar{\mathcal{C}}(\vec{y})\right]\neq 0.
	\end{equation}

To summarize, in our theory there are 21 constraints which can be divided into two classes:
\begin{equation}
\begin{aligned}7\;\text{first-class:} & \quad\pi_{i},\;\mathcal{C}_{i},\;\bar{p},\\
14\;\text{second-class:} & \quad p^{ij},\;\bar{\pi},\;\tilde{\pi}^{ij},\;\bar{\mathcal{C}}.
\end{aligned}
\end{equation}
The number of physical degrees of freedom is calculated to be 
	\begin{eqnarray}
	\#_{\mathrm{dof}} & = & \frac{1}{2}\left(2\times\#_{\mathrm{var}}-2\times\#_{\mathrm{1st}}-\#_{\mathrm{2nd}}\right)\nonumber \\
	& = & \frac{1}{2}\left(2\times17-2\times7-14\right)\nonumber \\
	& = & 3. \label{num_dof_3}
	\end{eqnarray}
We thus have shown that as long as the degeneracy condition (\ref{degen}) and consistency condition (\ref{calF_cond}) are satisfied, our theory has at most 3 physical degrees of freedom.

\section{Illustrating example: the quadratic case} \label{sec:exa}

According to the previous analysis, the degeneracy condition (\ref{degen}) by itself is merely a necessary condition to evade the unwanted scalar mode in general.
To fully remove the unwanted mode, the consistency condition (\ref{calF_cond}) must be imposed in order to guarantee the existence of the associated secondary constraint.
To see how this could happen and as an application of the formalism we developed,
let us consider the following Lagrangian:
	\begin{equation}
	\mathcal{L}^{\mathrm{(quad)}}=a_{1}K+a_{2}F+b_{1}K_{ij}K^{ij}+b_{2}K^{2}+c_{1}KF+c_{2}F^{2}+\mathcal{V}, \label{L_quad_ori}
	\end{equation}
where $K_{ij}$ is the extrinsic curvature and $F$ is given in  (\ref{F_Kij}), $\mathcal{V}$ is given by\footnote{Actually $\mathcal{V}$ can be chosen quite arbitrarily without changing the following analysis.}
	\begin{equation}
	\mathcal{V}=d_{1}+d_{2}R+d_{3}\nabla_{i}N\nabla^{i}N. \label{calV}
	\end{equation}
In the above all the coefficients $a_1$ etc. are generally functions of $t$ and $N$ and its first-order spatial derivatives, i.e., 
	\begin{equation}
		a_{1} = a_{1}(t,N,X),\qquad \text{with}\quad X\equiv \frac{1}{2}\partial_{i}N \partial^{i}N,
	\end{equation}
etc.
As we shall see, the dependence of the coefficient on $\partial_{i}N$ is crucial.

Generally, the theory described by $\mathcal{L}^{\mathrm{(quad)}}$ in (\ref{L_quad_ori}) propagates 4 physical degrees of freedom.
However, as we shall show below, it is possible to remove the unwanted mode by tuning the coefficients such that the degeneracy condition (\ref{degen}) and consistency condition (\ref{calF_cond}) are satisfied.
According to (\ref{S_fin}) and (\ref{L_tld_fin}), we write
	\begin{equation}
	S\equiv\int\mathrm{d}t\mathrm{d}^{3}x\,N\sqrt{h}\left(a_{1}B+a_{2}A+b_{1}B_{ij}B^{ij}+b_{2}B^{2}+c_{1}BA+c_{2}A^{2}+\mathcal{V}\right),
	\end{equation}
and
	\begin{equation}
	\tilde{S}\equiv S+\int\mathrm{d}t\mathrm{d}^{3}x\left[\frac{\delta S}{\delta A}\left(F-A\right)+\frac{\delta S}{\delta B_{ij}}\left(K_{ij}-B_{ij}\right)\right],
	\end{equation}
with
	\begin{eqnarray}
	\frac{\delta S}{\delta A} & = & N\sqrt{h}\left(a_{2}+c_{1}B+2c_{2}A\right),\\
	\frac{\delta S}{\delta B_{ij}} & = & N\sqrt{h}\left(a_{1}h^{ij}+2b_{1}B^{ij}+2b_{2}B\,h^{ij}+c_{1}Ah^{ij}\right).
	\end{eqnarray}
In our case,
	\begin{eqnarray}
	\frac{\delta^{2}S}{\delta A\left(\vec{x}\right)\delta A\left(\vec{y}\right)} & = & \delta^{3}\left(\vec{x}-\vec{y}\right) N\sqrt{h}2c_{2},\\
	\frac{\delta^{2}S}{\delta B_{ij}\left(\vec{x}\right)\delta A\left(\vec{y}\right)} & = & \delta^{3}\left(\vec{x}-\vec{y}\right) N\sqrt{h}c_{1}h^{ij},\\
	\frac{\delta^{2}S}{\delta B_{ij}\left(\vec{x}\right)\delta B_{kl}\left(\vec{y}\right)} & = & \delta^{3}\left(\vec{x}-\vec{y}\right) N\sqrt{h}\left[b_{1}\left(h^{ik}h^{jl}+h^{il}h^{jk}\right)+2b_{2}h^{ij}h^{kl}\right].
	\end{eqnarray}
The ``inverse'' defined as in (\ref{calG_fd2_BB}) can be got easily
	\begin{equation}
	\mathcal{G}_{ij,kl}\left(\vec{x},\vec{y}\right)=\delta^{3}\left(\vec{x}-\vec{y}\right)\frac{1}{N\sqrt{h}}\frac{1}{2b_{1}}\left[\frac{1}{2}\left(h_{ik}h_{jl}+h_{il}h_{jk}\right)-\frac{b_{2}}{b_{1}+3b_{2}}h_{ij}h_{kl}\right],
	\end{equation}
for $b_{1}\neq 0$ and $b_{1}+3b_{2}\neq 0$.

The degeneracy condition (\ref{degen}) thus reads
	\begin{equation}
	\mathcal{D}\left(\vec{x},\vec{y}\right)=\delta^{3}\left(\vec{x}-\vec{y}\right)N\sqrt{h}2\left(c_{2}-\frac{3}{4}\frac{c_{1}^{2}}{b_{1}+3b_{2}}\right)\equiv 0,
	\end{equation}
which implies we have to require
	\begin{equation}
		c_{2}=\frac{3}{4}\frac{c_{1}^{2}}{b_{1}+3b_{2}}. \label{exm_c2}
	\end{equation}
For later convenience, from now on we denote
\begin{equation}
b_{1}+3b_{2}\equiv 3\beta,\qquad c_{1}\equiv 2\beta\gamma,
\end{equation}
with $\beta \neq 0$.

The consistency condition (\ref{calF_cond}) is much more involved. After some manipulations, we find that we must require
	\begin{equation}
	\mathcal{F}(\vec{x},\vec{y})=\partial_{y^{i}}\delta^{3}(\vec{x}-\vec{y})\partial^{i}N(\vec{x})\sqrt{h(\vec{x})}\mathcal{E}(\vec{x})-\partial_{x^{i}}\delta^{3}(\vec{x}-\vec{y})\partial^{i}N(\vec{y})\sqrt{h(\vec{y})}\mathcal{E}(\vec{y}) \equiv 0, \label{calF_exm}
	\end{equation}
where we define
	\begin{equation}
	\mathcal{E}=\frac{\partial a_{2}}{\partial X}-\gamma\frac{\partial a_{1}}{\partial X}+2\left(B+A\gamma\right)\beta\frac{\partial\gamma}{\partial X}, \label{calE_def}
	\end{equation}
for short. 

Here comes the crucial point. In the special case where all the coefficients have no functional dependence on $\partial_{i}N$, we have $\frac{\partial f}{\partial X} \equiv 0$ with $f=a_1,a_2,\cdots$. In this case $\mathcal{E}\equiv  0$ and thus the consistency condition (\ref{calF_exm}) is automatically satisfied. This means the Lagrangian (\ref{L_quad_ori}) satisfying the degeneracy condition (\ref{exm_c2}), which now reads
	\begin{equation}
	\mathcal{L}^{\mathrm{(quad)}}=a_{1}K+a_{2}F+b_{1}\left(K_{ij}K^{ij}-\frac{1}{3}K^{2}\right)+\beta\left(K+\gamma F\right)^{2}+\mathcal{V},\label{L_quad_dege}
	\end{equation}
with $\mathcal{V}$ given in (\ref{calV}) and all the coefficients are functions of only $t$ and $N$, describes a class of healthy theories that propagate 3 physical degrees of freedom. 
At this point, we note that the model in \cite{Domenech:2015tca} got by a disformal transformation is a special case of (\ref{L_quad_dege}).

On the other hand, generally coefficients $a_{1}$ etc. can  depend on $\partial_{i}N$, and thus $\mathcal{E} \neq 0$ in general. In this case, although the kinetic terms are degenerate, the theory still possesses more than 3 degrees of freedom. Thus our example explicitly shows that the degeneracy condition is merely a necessary condition to evade the unwanted mode in general, due to the lack of an associated secondary constraint. In order to fully remove the unwanted mode, the consistency condition (\ref{calF_cond}) which now is equivalent to $\mathcal{E} = 0$ with $\mathcal{E}$ given in (\ref{calE_def}) must be imposed. This yields 2 differential equations for the coefficients:
	\begin{equation}
	\frac{\partial a_{2}}{\partial X}-\gamma\frac{\partial a_{1}}{\partial X}=0,\qquad\frac{\partial\gamma}{\partial X}=0,
	\end{equation}
which imply that $\gamma$ has no dependence on $X$, i.e.,
	\begin{equation}
	\gamma =\gamma\left(t,N\right),
	\end{equation}
and $a_1$ and $a_2$ must be related through
	\begin{equation}
		a_{2}-\gamma\left(t,N\right)a_{1} = \alpha\left(t,N\right),
	\end{equation}
with $\alpha$ some general function of $t$ and $N$.
Finally, the Lagrangian satisfying both the degeneracy and consistency conditions reads
	\begin{equation}
	\mathcal{L}^{\mathrm{(quad)}}=a_{1}\left(K+\gamma F\right)+\alpha F+b_{1}\left(K_{ij}K^{ij}-\frac{1}{3}K^{2}\right)+\beta\left(K+\gamma F\right)^{2}+\mathcal{V}.\label{L_quad_fin}
	\end{equation}
We thus get a class of theories of which the kinetic terms
(strictly speaking, terms involving temporal derivatives) are controlled by 5 coefficients
$a_{1},b_{1},\alpha,\beta,\gamma$ (with $b_1,\beta \neq 0$), where $a_{1},b_{1},\beta$ can be generally functions of $t,N,\partial_{i}N$, while $\alpha,\gamma$  must be functions of $t$ and $N$ only. As for coefficients in the potential terms $\mathcal{V}$, there is no
restriction. It is clear that
(\ref{L_quad_fin}) includes (\ref{L_quad_dege}) as a special case.

\section{Conclusion} \label{sec:concl}

In this work, we investigated the extension to the framework proposed in \cite{Gao:2014soa} by including the velocity of the lapse function $\dot{N}$ as one of the basic ingredients.
From the geometric point of view, $\dot{N}$ (in terms of $\pounds_{\bm{n}}N$) should be treated in the same footing as $\dot{h}_{ij}$ (in terms of $\pounds_{\bm{n}}h_{ij} \equiv 2K_{ij}$).
Thus our general Lagrangian given in (\ref{S_ori}) is not only a natural extension to that in \cite{Gao:2014soa}, but also provides a more complete framework for spatially covariant gravity theories.

Contrary to the framework in \cite{Gao:2014soa} which is always ``safe'', generally the Lagrangian (\ref{S_ori}) will have 4 physical degrees of freedom, one of which is the unwanted scalar mode.
Nevertheless, through a detailed Hamiltonian analysis, we find two conditions to prevent this unwanted degree of freedom.
One is the degeneracy condition (\ref{degen}) which is essentially the requirement of the degeneracy of the kinetic terms in the original Lagrangian (\ref{S_ori}).
The other is the consistency condition (\ref{calF_cond}), which ensures the existence of an additional secondary constraint $\mathcal{C}$ given in (\ref{calC_def}).
These two conditions will restrict the structure of our general Lagrangian (\ref{S_ori}).
As long as these two conditions are satisfied, there are at most 3 physical degrees of freedom are propagating.

According to our analysis, the degeneracy condition is merely a necessary condition in general, and thus by itself is not sufficient to remove the unwanted mode.
This is because in general the primary constraint due to the degeneracy condition does not necessarily induce a secondary constraint. 
For the theories considered in this work, this generally happens when there are mixings between the temporal and spatial derivatives in the Lagrangian.
These include terms such as $f(t,N,\partial_{i}N,\cdots) K$ etc., which are consider in the example in Sec.\ref{sec:exa}, as well as terms such as $\nabla_{i}K \nabla^{i}K$, $\nabla_{i}F\nabla^{i}F$ etc.
In general, multiple conditions are required to reduce a whole degree of freedom \cite{Henneaux:1992ig,Langlois:2015skt,Motohashi:2016ftl,Crisostomi:2017aim,Deffayet:2015qwa,Motohashi:2014opa,Motohashi:2017eya,Motohashi:2018pxg,Motohashi:2016prk}.  Here in this paper our framework provides another example in which a single degeneracy condition is not sufficient.

Comments are in order.
Firstly, it has been shown that some of the theories in \cite{Langlois:2015cwa} can be reduced to Horndeski theory via disformal transformation \cite{BenAchour:2016fzp}, while the theories within the framework developed in \cite{Gao:2014soa} do contain subsets that cannot be transformed from Horndeski theory under the usual disformal transformation \cite{Fujita:2015ymn}.
It is thus useful to perform a systematic investigation of the property of our theory under general field transformations in order to identify the genuinely new and independent theories.
Secondly, as a first attempt, in this work we only consider the velocities of $h_{ij}$ and $N$, i.e., their first-order time derivatives. It would be interesting to consider higher-order time derivatives following the approach discussed in \cite{Motohashi:2017eya,Motohashi:2018pxg}.
As a final remark, we emphasize that the theories constructed in our paper respect only spatial symmetries, which are related to covariant theories in the unitary gauge. If one focuses on the degeneracy of the theories, one may conclude that the unitary gauge is misleading since non-degenerate theories appear to be degenerate in the unitary gauge \cite{Langlois:2015cwa,Langlois:2015skt,Crisostomi:2016tcp}, although the gauge fixing itself is legitimate \cite{Motohashi:2016prk}. 
Indeed, the generally covariant version of the spatially covariant gravity contains higher-order time derivatives both in the Lagrangian and in the equations of motion, which indicates the existence of unwanted degree(s) of freedom.
On the other hand, it was argued recently in \cite{DeFelice:2018ewo} (see \cite{Blas:2009yd} for an early work on this point) that this apparently dangerous mode can be made non-dynamical by choosing appropriate boundary conditions.
We will go back to these issues in the future.

\acknowledgments

We would like to thank Rio Saitou for useful correspondence. XG would like to thank Masahide Yamaguchi and Jun'ichi Yokoyama for discussion and for hospitality during his visit in Tokyo.
This work was supported by the Chinese National
Youth Thousand Talents Program (No. 71000-41180003) and by the SYSU start-up funding (No. 71000-52601106).

\appendix

\section{Lie derivatives of tensor densities} \label{sec:Lder_td}

For a general spatial tensor density $\bm{Y}$ of weight unity (i.e., $\bm{Y}/\sqrt{h}$ behaves as a spatial tensor field), its Lie derivative with respect to an arbitrary spatial vector field $\vec{\xi}$ is defined by
	\begin{equation}
		\pounds_{\vec{\xi}}\, \bm{Y} := \delta_{\vec{\xi}} \, \bm{Y},
	\end{equation}
where $\delta_{\vec{\xi}} \, \bm{Y}$ denotes the infinitesimal change under the spatial diffeomorphism generated by $\vec{\xi}$. 
Since $\bm{Y}/\sqrt{h}$ behaves as a spatial tensor field, we have
\begin{equation}
\delta_{\vec{\xi}}\left(\frac{\bm{Y}}{\sqrt{h}}\right)=\pounds_{\vec{\xi}}\left(\frac{\bm{Y}}{\sqrt{h}}\right).
\end{equation}
The left-hand-side can be expanded straightforwardly as
\begin{eqnarray}
\delta_{\vec{\xi}}\left(\frac{\bm{Y}}{\sqrt{h}}\right) & = & \frac{1}{\sqrt{h}}\delta_{\vec{\xi}}\,\bm{Y}+\bm{Y}\,\delta_{\vec{\xi}}\left(\frac{1}{\sqrt{h}}\right)\nonumber \\
& = & \frac{1}{\sqrt{h}}\delta_{\vec{\xi}}\,\bm{Y}-\frac{1}{\sqrt{h}}\frac{1}{2}\bm{Y}\,\delta_{\vec{\xi}}\ln h\nonumber \\
& = & \frac{1}{\sqrt{h}}\delta_{\vec{\xi}}\,\bm{Y}-\frac{1}{\sqrt{h}}\frac{1}{2}\bm{Y}\,h^{ij}\delta_{\vec{\xi}}\,h_{ij},
\end{eqnarray}
then using $\delta_{\vec{\xi}}\,h_{ij}\equiv\pounds_{\vec{\xi}}\,h_{ij}$,
we get
\begin{equation}
\pounds_{\vec{\xi}}\,\bm{Y}\equiv\delta_{\vec{\xi}}\,\bm{Y}=\sqrt{h}\pounds_{\vec{\xi}}\left(\frac{\bm{Y}}{\sqrt{h}}\right)+\frac{1}{2}\bm{Y}\,h^{ij}\pounds_{\vec{\xi}}\,h_{ij}.\label{del_xi_td_gen}
\end{equation}
It is convenient to derive the corresponding expression for the components.
Supposing $\bm{Y}/\sqrt{h}$ is a $(m,n)$-type spatial tensor field, simple manipulations yield
	\begin{eqnarray}
	\pounds_{\vec{\xi}}\,Y_{\phantom{i_{1}\cdots i_{m}}j_{1}\cdots j_{n}}^{i_{1}\cdots i_{m}} & = & \partial_{k}\left(\xi^{k}Y_{\phantom{i_{1}\cdots i_{m}}j_{1}\cdots j_{n}}^{i_{1}\cdots i_{m}}\right)\nonumber \\
	&  & -\sum_{\ell=1}^{m}Y_{\phantom{i_{1}\cdots k\cdots i_{m}}j_{1}\cdots j_{n}}^{i_{1}\cdots k\cdots i_{m}}\partial_{k}\xi^{i_{\ell}}+\sum_{\ell=1}^{n}Y_{\phantom{i_{1}\cdots i_{m}}j_{1}\cdots k\cdots j_{n}}^{i_{1}\cdots i_{m}}\partial_{j_{\ell}}\xi^{k}.\label{Lder_td}
	\end{eqnarray}
In terms of spatial covariant derivatives, we have
	\begin{eqnarray}
	\pounds_{\vec{\xi}}\,Y_{\phantom{i_{1}\cdots i_{m}}j_{1}\cdots j_{n}}^{i_{1}\cdots i_{m}} & = & \sqrt{h}\nabla_{k}\left(\xi^{k}\frac{Y_{\phantom{i_{1}\cdots i_{m}}j_{1}\cdots j_{n}}^{i_{1}\cdots i_{m}}}{\sqrt{h}}\right)\nonumber \\
	&  & -\sum_{\ell=1}^{m}Y_{\phantom{i_{1}\cdots k\cdots i_{m}}j_{1}\cdots j_{n}}^{i_{1}\cdots k\cdots i_{m}}\nabla_{k}\xi^{i_{\ell}}+\sum_{\ell=1}^{n}Y_{\phantom{i_{1}\cdots i_{m}}j_{1}\cdots k\cdots j_{n}}^{i_{1}\cdots i_{m}}\nabla_{j_{\ell}}\xi^{k}.\label{Lder_td_cov}
	\end{eqnarray}

In our case, (\ref{Lder_td}) or equivalently (\ref{Lder_td_cov})
immediately yields
	\begin{equation}
	\pounds_{\vec{\xi}}\,\pi=\partial_{i}\left(\xi^{i}\,\pi\right), \label{Lder_pi}
	\end{equation}
which is a total derivative, and
	\begin{equation}
	\pounds_{\vec{\xi}}\,\pi^{ij} = \sqrt{h}\nabla_{k}\left(\xi^{k}\frac{\pi^{ij}}{\sqrt{h}}\right)-2\nabla_{k}\xi^{(i}\,\pi^{j)k}. \label{Lder_pi^ij}
	\end{equation}

\section{Proof of (\ref{PB_LD_xi})} \label{app:PB_LD}

In this appendix, we prove (\ref{PB_LD_xi}) with a more general setting.

Supposing at each local point in space, the phase space is spanned by variables $\left\{ \phi_{I},\pi^{I}\right\}$ (not necessarily the variables $\{\Phi_{I},\Pi^{I}\}$ in the main text), with $\phi$'s the canonical variables and $\pi$'s the conjugate momenta.
We assume that $\phi$'s are spatial tensor fields and thus $\pi$'s are spatial tensor densities.
Here the index $I$ formally denotes different kinds of variables as well as their tensorial indices.

We define a general functional $X[\vec{\xi}]$ by
	\begin{equation}
	X[\vec{\xi}]:=\int\mathrm{d}^{3}x\sum_{I}\pi^{I}\pounds_{\vec{\xi}}\,\phi_{I},\label{Xfn_def_gen}
	\end{equation}
where the summation runs over \emph{all} the variables, $\vec{\xi}$ is an arbitrary spatial vector field that may or may not depend on the phase space variables. 
For any functional on the phase space $\mathcal{F} = \mathcal{F}[\phi_{I},\pi^{I};q_{a}]$ that is invariant under the time-independent spatial diffeomorphism, where $\{q_{a}\}$ stands for variables independent of $\{\phi_{I},\pi^{I}\}$,
we will show that the Poisson bracket of $X[\vec{\xi}]$ with $\mathcal{F}$ reads
	\begin{equation}
	\left[\int\mathrm{d}^{3}x\sum_{I}\pi^{I}\pounds_{\vec{\xi}}\,\phi_{I},\mathcal{F}\right]=\int\mathrm{d}^{3}x\sum_{I}\pi^{I}\pounds_{[\vec{\xi},\mathcal{F}]}\phi_{I} +\int\mathrm{d}^{3}x\sum_{a}\frac{\delta\mathcal{F}}{\delta q_{a}}\pounds_{\vec{\xi}}\,q_{a},\label{PB_LD_gen}
	\end{equation}
or compactly
	\begin{equation}
	\left[X[\vec{\xi}],\mathcal{F}\right]=X\left[[\vec{\xi},\mathcal{F}]\right] +\int\mathrm{d}^{3}x\sum_{a}\frac{\delta\mathcal{F}}{\delta q_{a}}\pounds_{\vec{\xi}}\,q_{a},\label{PB_LD_com}
	\end{equation}
up to a boundary term.
Please note since the Lie derivative of a scalar density is a total derivative (see (\ref{Lder_pi}))
	\begin{equation}
	\sum_{I}\pounds_{\vec{\xi}}\,\left(\pi^{I}\phi_{I}\right)\equiv\sum_{I}\pi^{I}\pounds_{\vec{\xi}}\,\phi_{I}+\sum_{I}\phi_{I}\pounds_{\vec{\xi}}\,\pi^{I}\equiv\partial_{i}\left(\xi^{i}\sum_{I}\pi^{I}\phi_{I}\right),
	\end{equation}
$X[\vec{\xi}]$ defined in (\ref{Xfn_def_gen}) can be written equivalently as
	\begin{equation}
	X[\vec{\xi}]\simeq-\int\mathrm{d}^{3}x\sum_{I}\phi_{I}\pounds_{\vec{\xi}}\,\pi^{I}.
	\end{equation}

Before proving (\ref{PB_LD_gen}), note for an arbitrary spatial tensor field  $\bm{T}$, the following identity holds:
	\begin{equation}
	\left[\pounds_{\vec{\xi}}\,\bm{T},\mathcal{F}\right]=\pounds_{\left[\vec{\xi},\mathcal{F}\right]}\bm{T}+\pounds_{\vec{\xi}}\left[\bm{T},\mathcal{F}\right],\label{PB_LD}
	\end{equation}
where again $\vec{\xi}$ is an arbitrary spatial vector, $\mathcal{F}$ is an arbitrary scalar functional.
This simply follows from the fact that both Poisson bracket with a scalar functional $\left[\,\cdot\,,\mathcal{F}\right]$ and Lie derivative $\pounds_{\vec{\xi}}$ are linear derivative operators obeying the Leibniz rule, and both preserve the types of tensor fields they act on.
In fact, (\ref{PB_LD}) can be verified explicitly by plugging the expression for the components of $\pounds_{\vec{\xi}}\,\bm{T}$. Supposing $\bm{T}$ is a $(m,n)$-type spatial tensor fields, we have
	\begin{eqnarray}
	&  & \left[\pounds_{\vec{\xi}}\,T_{\phantom{i_{1}\cdots i_{m}}j_{1}\cdots j_{n}}^{i_{1}\cdots i_{m}},\mathcal{F}\right]\nonumber \\
	& = & \left[\xi^{k}\partial_{k}T_{\phantom{i_{1}\cdots i_{p}}j_{1}\cdots j_{q}}^{i_{1}\cdots i_{p}}-\sum_{\ell=1}^{m}T_{\phantom{i_{1}\cdots k\cdots i_{m}}j_{1}\cdots j_{n}}^{i_{1}\cdots k\cdots i_{m}}\partial_{k}\xi^{i_{\ell}}+\sum_{\ell=1}^{n}T_{\phantom{i_{1}\cdots i_{m}}j_{1}\cdots k\cdots j_{n}}^{i_{1}\cdots i_{m}}\partial_{j_{\ell}}\xi^{k},\mathcal{F}\right] \nonumber\\
	& = & \left[\xi^{k},\mathcal{F}\right]\partial_{k}T_{\phantom{i_{1}\cdots i_{p}}j_{1}\cdots j_{q}}^{i_{1}\cdots i_{p}}-\sum_{\ell=1}^{m}T_{\phantom{i_{1}\cdots k\cdots i_{m}}j_{1}\cdots j_{n}}^{i_{1}\cdots k\cdots i_{m}}\partial_{k}\left[\xi^{i_{\ell}},\mathcal{F}\right]+\sum_{\ell=1}^{n}T_{\phantom{i_{1}\cdots i_{m}}j_{1}\cdots k\cdots j_{n}}^{i_{1}\cdots i_{m}}\partial_{j_{\ell}}\left[\xi^{k},\mathcal{F}\right]\nonumber \\
	&  & +\xi^{k}\partial_{k}\left[T_{\phantom{i_{1}\cdots i_{p}}j_{1}\cdots j_{q}}^{i_{1}\cdots i_{p}},\mathcal{F}\right]-\sum_{\ell=1}^{m}\left[T_{\phantom{i_{1}\cdots k\cdots i_{m}}j_{1}\cdots j_{n}}^{i_{1}\cdots k\cdots i_{m}},\mathcal{F}\right]\partial_{k}\xi^{i_{\ell}}+\sum_{\ell=1}^{n}\left[T_{\phantom{i_{1}\cdots i_{m}}j_{1}\cdots k\cdots j_{n}}^{i_{1}\cdots i_{m}},\mathcal{F}\right]\partial_{j_{\ell}}\xi^{k}\nonumber \\
	& \equiv & \pounds_{\left[\vec{\xi},\mathcal{F}\right]}\,T_{\phantom{i_{1}\cdots i_{p}}j_{1}\cdots j_{q}}^{i_{1}\cdots i_{p}}+\pounds_{\vec{\xi}}\,\left[T_{\phantom{i_{1}\cdots i_{m}}j_{1}\cdots j_{n}}^{i_{1}\cdots i_{m}},\mathcal{F}\right],
	\end{eqnarray}
which proves (\ref{PB_LD}).
For later convenience, we also note 
	\begin{equation}
	\delta_{\vec{\xi}}\,\mathcal{F}=\int\mathrm{d}^{3}x\left[\sum_{I}\left(\frac{\delta\mathcal{F}}{\delta\phi_{I}}\pounds_{\vec{\xi}}\,\phi_{I}+\frac{\delta\mathcal{F}}{\delta\pi^{I}}\pounds_{\vec{\xi}}\,\pi^{I}\right)+\sum_{a}\frac{\delta\mathcal{F}}{\delta q_{a}}\pounds_{\vec{\xi}}\,q_{a}\right]\equiv 0, \label{delta_calF}
	\end{equation}
which follows from the fact that $\mathcal{F}$ is a scalar functional that is invariant under time-independent spatial diffeomorphism. In (\ref{delta_calF}) the Lie derivatives of tensor densities $\pounds_{\vec{\xi}}\,\pi^{I}$ are defined as in Sec.\ref{sec:Lder_td}.

Now we turn to the proof of (\ref{PB_LD_gen}).
Using (\ref{PB_LD}), the left-hand-side can be evaluated as
	\begin{eqnarray}
	\mathrm{L.H.S.} & \equiv & \left[\int\mathrm{d}^{3}x\sum_{I}\pi^{I}\pounds_{\vec{\xi}}\,\phi_{I},\mathcal{F}\right]\nonumber \\
	& = & \int\mathrm{d}^{3}x\sum_{I}\left[\pi^{I},\mathcal{F}\right]\pounds_{\vec{\xi}}\,\phi_{I}+\int\mathrm{d}^{3}x\sum_{I}\pi^{I}\left[\pounds_{\vec{\xi}}\,\phi_{I},\mathcal{F}\right]\nonumber \\
	& = & \int\mathrm{d}^{3}x\sum_{I}\left(-\frac{\delta\mathcal{F}}{\delta\phi_{I}}\right)\pounds_{\vec{\xi}}\,\phi_{I}+\int\mathrm{d}^{3}x\sum_{I}\pi^{I}\pounds_{\vec{\xi}}\left[\phi_{I},\mathcal{F}\right]+\int\mathrm{d}^{3}x\sum_{I}\pi^{I}\pounds_{\left[\vec{\xi},\mathcal{F}\right]}\phi_{I}\nonumber \\
	& = & -\int\mathrm{d}^{3}x\sum_{I}\frac{\delta\mathcal{F}}{\delta\phi_{I}}\pounds_{\vec{\xi}}\,\phi_{I}+\int\mathrm{d}^{3}x\sum_{I}\pi^{I}\pounds_{\vec{\xi}}\,\frac{\delta\mathcal{F}}{\delta\pi^{I}}+\int\mathrm{d}^{3}x\sum_{I}\pi^{I}\pounds_{\left[\vec{\xi},\mathcal{F}\right]}\phi_{I}. \label{PB_lhs_t1}
	\end{eqnarray}
Since $\sum_{I}\pi^{I}\frac{\delta\mathcal{F}}{\delta\pi^{I}}$ is a scalar density, its Lie derivative obeys that in (\ref{Lder_pi}),
	i.e.,
	\begin{equation}
	\int\mathrm{d}^{3}x\pounds_{\vec{\xi}}\left(\sum_{I}\pi^{I}\frac{\delta\mathcal{F}}{\delta\pi^{I}}\right)=\int\mathrm{d}^{3}x\,\partial_{i}\left(\xi^{i}\sum_{I}\pi^{I}\frac{\delta\mathcal{F}}{\delta\pi^{I}}\right)\simeq0,
	\end{equation}
which implies
	\begin{equation}
	\int\mathrm{d}^{3}x\sum_{I}\pi^{I}\pounds_{\vec{\xi}}\,\frac{\delta\mathcal{F}}{\delta\pi^{I}}\simeq-\int\mathrm{d}^{3}x\sum_{I}\frac{\delta\mathcal{F}}{\delta\pi^{I}}\pounds_{\vec{\xi}}\,\pi^{I},
	\end{equation}
where the Lie derivative of tensor densities $\pi^{I}$ are defined as in Sec.\ref{sec:Lder_td}. Plugging the above into (\ref{PB_lhs_t1}), we get
	\begin{eqnarray}
	\mathrm{L.H.S.} & \simeq & -\int\mathrm{d}^{3}x\sum_{I}\frac{\delta\mathcal{F}}{\delta\phi_{I}}\pounds_{\vec{\xi}}\,\phi_{I}-\int\mathrm{d}^{3}x\sum_{I}\frac{\delta\mathcal{F}}{\delta\pi^{I}}\pounds_{\vec{\xi}}\,\pi^{I}+\int\mathrm{d}^{3}x\sum_{I}\pi^{I}\pounds_{\left[\vec{\xi},\mathcal{F}\right]}\phi_{I}\nonumber \\
	& = & -\int\mathrm{d}^{3}x\sum_{I}\left(\frac{\delta\mathcal{F}}{\delta\phi_{I}}\pounds_{\vec{\xi}}\,\phi_{I}+\frac{\delta\mathcal{F}}{\delta\pi^{I}}\pounds_{\vec{\xi}}\,\pi^{I}\right)+\int\mathrm{d}^{3}x\sum_{I}\pi^{I}\pounds_{\left[\vec{\xi},\mathcal{F}\right]}\phi_{I}\nonumber \\
	& \equiv & \int\mathrm{d}^{3}x\sum_{I}\pi^{I}\pounds_{\left[\vec{\xi},\mathcal{F}\right]}\phi_{I} +\int\mathrm{d}^{3}x\sum_{a}\frac{\delta\mathcal{F}}{\delta q_{a}}\pounds_{\vec{\xi}}\,q_{a},
	\end{eqnarray}
which is just (\ref{PB_LD_gen}). 
In the last step, we used (\ref{delta_calF}), i.e. $\delta_{\vec{\xi}}\,\mathcal{F}\equiv 0$, which follows from the fact that $\mathcal{F}$ is a scalar functional that is invariant under time-independent spatial diffeomorphism.

\section{Poisson brackets for the primary constraints} \label{app:PB_pc}

In this appendix, we present some details in calculating the Poisson brackets for the primary constraints.

We first derive some general results for later convenience. For a
general phase space function or functional $F$, we have
	\begin{eqnarray}
	\left[\pi_{i}\left(\vec{x}\right),F\right] & = & -\frac{\delta F}{\delta N^{i}\left(\vec{x}\right)},\label{PB_pi_i_gen}\\
	\left[p\left(\vec{x}\right),F\right] & = & -\frac{\delta F}{\delta A\left(\vec{x}\right)},\label{PB_p_gen}\\
	\left[p^{ij}\left(\vec{x}\right),F\right] & = & -\frac{\delta F}{\delta B_{ij}\left(\vec{x}\right)}.\label{PB_p^ij_gen}
	\end{eqnarray}

Recalling that $\tilde{\pi}$ is defined in (\ref{pri_cons_2}) and using the fact that $S$ in (\ref{S_fin}) generally depends
on $N,h_{ij},A,B_{ij}$ but not on $N_{i}$, we get
	\begin{eqnarray}
	\left[\tilde{\pi}\left(\vec{x}\right),F\right] & = & -\frac{\delta F}{\delta N\left(\vec{x}\right)}+\int\mathrm{d}^{3}z\bigg[\frac{\delta\tilde{\pi}\left(\vec{x}\right)}{\delta N\left(\vec{z}\right)}\frac{\delta F}{\delta\pi\left(\vec{z}\right)}+\frac{\delta\tilde{\pi}\left(\vec{x}\right)}{\delta h_{ij}\left(\vec{z}\right)}\frac{\delta F}{\delta\pi^{ij}\left(\vec{z}\right)}\nonumber \\
	&  & \quad+\frac{\delta\tilde{\pi}\left(\vec{x}\right)}{\delta A\left(\vec{z}\right)}\frac{\delta F}{\delta p\left(\vec{z}\right)}+\frac{\delta\tilde{\pi}\left(\vec{x}\right)}{\delta B_{ij}\left(\vec{z}\right)}\frac{\delta F}{\delta p^{ij}\left(\vec{z}\right)}\bigg].\label{PB_pitld_ori}
	\end{eqnarray}
Various functional
derivatives of $\tilde{\pi}$ are calculated to be
	\begin{eqnarray}
	\frac{\delta\tilde{\pi}\left(\vec{x}\right)}{\delta N\left(\vec{z}\right)} & = & \delta^{3}\left(\vec{x}-\vec{z}\right)\frac{1}{N^{2}\left(\vec{x}\right)}\frac{\delta S}{\delta A\left(\vec{x}\right)}-\frac{1}{N\left(\vec{x}\right)}\frac{\delta^{2}S}{\delta A\left(\vec{x}\right)\delta N\left(\vec{z}\right)},\label{fd_pitld_N}\\
	\frac{\delta\tilde{\pi}\left(\vec{x}\right)}{\delta h_{ij}\left(\vec{z}\right)} & = & -\frac{1}{N\left(\vec{x}\right)}\frac{\delta^{2}S}{\delta A\left(\vec{x}\right)\delta h_{ij}\left(\vec{z}\right)},\label{fd_pitld_hij}\\
	\frac{\delta\tilde{\pi}\left(\vec{x}\right)}{\delta A\left(\vec{z}\right)} & = & -\frac{1}{N\left(\vec{x}\right)}\frac{\delta^{2}S}{\delta A\left(\vec{x}\right)\delta A\left(\vec{z}\right)},\label{fd_pitld_A}\\
	\frac{\delta\tilde{\pi}\left(\vec{x}\right)}{\delta B_{ij}\left(\vec{z}\right)} & = & -\frac{1}{N\left(\vec{x}\right)}\frac{\delta^{2}S}{\delta A\left(\vec{x}\right)\delta B_{ij}\left(\vec{z}\right)}.\label{fd_pitld_Bij}
	\end{eqnarray}
Plugging the above into (\ref{PB_pitld_ori}), we find
	\begin{eqnarray}
	\left[\tilde{\pi}\left(\vec{x}\right),F\right] & = & -\frac{\delta F}{\delta N\left(\vec{x}\right)}+\frac{1}{N^{2}\left(\vec{x}\right)}\frac{\delta S}{\delta A\left(\vec{x}\right)}\frac{\delta F}{\delta\pi\left(\vec{x}\right)}\nonumber \\
	&  & -\frac{1}{N\left(\vec{x}\right)}\int\mathrm{d}^{3}z\bigg[\frac{\delta^{2}S}{\delta A\left(\vec{x}\right)\delta N\left(\vec{z}\right)}\frac{\delta F}{\delta\pi\left(\vec{z}\right)}+\frac{\delta^{2}S}{\delta A\left(\vec{x}\right)\delta h_{ij}\left(\vec{z}\right)}\frac{\delta F}{\delta\pi^{ij}\left(\vec{z}\right)}\nonumber \\
	&  & \quad+\frac{\delta^{2}S}{\delta A\left(\vec{x}\right)\delta A\left(\vec{z}\right)}\frac{\delta F}{\delta p\left(\vec{z}\right)}+\frac{\delta^{2}S}{\delta A\left(\vec{x}\right)\delta B_{ij}\left(\vec{z}\right)}\frac{\delta F}{\delta p^{ij}\left(\vec{z}\right)}\bigg].\label{PB_pitld_gen}
	\end{eqnarray}

Similarly, for $\tilde{\pi}^{ij}$ defined in (\ref{pri_cons_2}), we have
	\begin{eqnarray}
	\left[\tilde{\pi}^{ij}\left(\vec{x}\right),F\right] & = & -\frac{\delta F}{\delta h_{ij}\left(\vec{x}\right)}+\int\mathrm{d}^{3}z\bigg[\frac{\delta\tilde{\pi}^{ij}\left(\vec{x}\right)}{\delta N\left(\vec{z}\right)}\frac{\delta F}{\delta\pi\left(\vec{z}\right)}+\frac{\delta\tilde{\pi}^{ij}\left(\vec{x}\right)}{\delta h_{kl}\left(\vec{z}\right)}\frac{\delta F}{\delta\pi^{kl}\left(\vec{z}\right)}\nonumber \\
	&  & \quad+\frac{\delta\tilde{\pi}^{ij}\left(\vec{x}\right)}{\delta A\left(\vec{z}\right)}\frac{\delta F}{\delta p\left(\vec{z}\right)}+\frac{\delta\tilde{\pi}^{ij}\left(\vec{x}\right)}{\delta B_{kl}\left(\vec{z}\right)}\frac{\delta F}{\delta p^{kl}\left(\vec{z}\right)}\bigg].\label{PB_pitld^ij_ori}
	\end{eqnarray}
where the functional derivatives are found to be
	\begin{eqnarray}
	\frac{\delta\tilde{\pi}^{ij}\left(\vec{x}\right)}{\delta N\left(\vec{z}\right)} & = & \frac{1}{2}\delta^{3}\left(\vec{x}-\vec{z}\right)\frac{1}{N^{2}\left(\vec{x}\right)}\frac{\delta S}{\delta B_{ij}\left(\vec{x}\right)}-\frac{1}{2}\frac{1}{N\left(\vec{x}\right)}\frac{\delta^{2}S}{\delta B_{ij}\left(\vec{x}\right)\delta N\left(\vec{z}\right)},\label{fd_pitld^ij_N}\\
	\frac{\delta\tilde{\pi}^{ij}\left(\vec{x}\right)}{\delta h_{kl}\left(\vec{z}\right)} & = & -\frac{1}{2N\left(\vec{x}\right)}\frac{\delta^{2}S}{\delta B_{ij}\left(\vec{x}\right)\delta h_{kl}\left(\vec{z}\right)},\label{fd_pitld^ij_hkl}\\
	\frac{\delta\tilde{\pi}^{ij}\left(\vec{x}\right)}{\delta A\left(\vec{z}\right)} & = & -\frac{1}{2N\left(\vec{x}\right)}\frac{\delta^{2}S}{\delta B_{ij}\left(\vec{x}\right)\delta A\left(\vec{z}\right)},\label{fd_pitld^ij_A}\\
	\frac{\delta\tilde{\pi}^{ij}\left(\vec{x}\right)}{\delta B_{kl}\left(\vec{z}\right)} & = & -\frac{1}{2N\left(\vec{x}\right)}\frac{\delta^{2}S}{\delta B_{ij}\left(\vec{x}\right)\delta B_{kl}\left(\vec{z}\right)}.\label{fd_pitld^ij_Bkl}
	\end{eqnarray}
Plugging the above into (\ref{PB_pitld^ij_ori}), we get
	\begin{eqnarray}
	\left[\tilde{\pi}^{ij}\left(\vec{x}\right),F\right] & = & -\frac{\delta F}{\delta h_{ij}\left(\vec{x}\right)}+\frac{1}{2}\frac{1}{N^{2}\left(\vec{x}\right)}\frac{\delta S}{\delta B_{ij}\left(\vec{x}\right)}\frac{\delta F}{\delta\pi\left(\vec{x}\right)}\nonumber \\
	&  & -\frac{1}{2}\frac{1}{N\left(\vec{x}\right)}\int\mathrm{d}^{3}z\bigg[\frac{\delta^{2}S}{\delta B_{ij}\left(\vec{x}\right)\delta N\left(\vec{z}\right)}\frac{\delta F}{\delta\pi\left(\vec{z}\right)}+\frac{\delta^{2}S}{\delta B_{ij}\left(\vec{x}\right)\delta h_{kl}\left(\vec{z}\right)}\frac{\delta F}{\delta\pi^{kl}\left(\vec{z}\right)}\nonumber \\
	&  & \quad+\frac{\delta^{2}S}{\delta B_{ij}\left(\vec{x}\right)\delta A\left(\vec{z}\right)}\frac{\delta F}{\delta p\left(\vec{z}\right)}+\frac{\delta^{2}S}{\delta B_{ij}\left(\vec{x}\right)\delta B_{kl}\left(\vec{z}\right)}\frac{\delta F}{\delta p^{kl}\left(\vec{z}\right)}\bigg].\label{PB_pitld^ij_gen}
	\end{eqnarray}

Now we evaluate the Poisson brackets among the primary constraints.

From (\ref{PB_pi_i_gen}), since all the primary constraints
$\left\{ \varphi^{\alpha}\right\} $ have no functional dependence
on $N^{i}$, we get
\begin{equation}
\left[\pi_{i}\left(\vec{x}\right),\varphi^{\alpha}\left(\vec{y}\right)\right]=-\frac{\delta\varphi^{\alpha}\left(\vec{y}\right)}{\delta N^{i}\left(\vec{x}\right)}\equiv 0.
\end{equation}

From (\ref{PB_p_gen}), we get
	\begin{eqnarray}
	\left[p\left(\vec{x}\right),p\left(\vec{y}\right)\right] & = & -\frac{\delta p\left(\vec{y}\right)}{\delta A\left(\vec{x}\right)}\equiv 0, \label{PB_p_p}\\
	\left[p\left(\vec{x}\right),p^{ij}\left(\vec{y}\right)\right] & = & -\frac{\delta p^{ij}\left(\vec{y}\right)}{\delta A\left(\vec{x}\right)}\equiv 0, \label{PB_p_p^ij}
	\end{eqnarray}
and
	\begin{eqnarray}
	\left[p\left(\vec{x}\right),\tilde{\pi}\left(\vec{y}\right)\right] & = & -\frac{\delta\tilde{\pi}\left(\vec{y}\right)}{\delta A\left(\vec{x}\right)}=\frac{1}{N\left(\vec{y}\right)}\frac{\delta^{2}S}{\delta A\left(\vec{x}\right)\delta A\left(\vec{y}\right)},\\
	\left[p\left(\vec{x}\right),\tilde{\pi}^{ij}\left(\vec{y}\right)\right] & = & -\frac{\delta\tilde{\pi}^{ij}\left(\vec{y}\right)}{\delta A\left(\vec{x}\right)}=\frac{1}{2N\left(\vec{y}\right)}\frac{\delta^{2}S}{\delta A\left(\vec{x}\right)\delta B_{ij}\left(\vec{y}\right)},
	\end{eqnarray}
where we used (\ref{fd_pitld_A}) and (\ref{fd_pitld^ij_A}).

From (\ref{PB_p^ij_gen}),
\begin{equation}
\left[p^{ij}\left(\vec{x}\right),p^{kl}\left(\vec{y}\right)\right]=-\frac{\delta p^{kl}\left(\vec{y}\right)}{\delta B_{ij}\left(\vec{x}\right)}\equiv0,
\end{equation}
and
\begin{eqnarray}
\left[p^{ij}\left(\vec{x}\right),\tilde{\pi}\left(\vec{y}\right)\right] & = & -\frac{\delta\tilde{\pi}\left(\vec{y}\right)}{\delta B_{ij}\left(\vec{x}\right)}=\frac{1}{N\left(\vec{y}\right)}\frac{\delta^{2}S}{\delta B_{ij}\left(\vec{x}\right)\delta A\left(\vec{y}\right)},\\
\left[p^{ij}\left(\vec{x}\right),\tilde{\pi}^{kl}\left(\vec{y}\right)\right] & = & -\frac{\delta\tilde{\pi}^{kl}\left(\vec{y}\right)}{\delta B_{ij}\left(\vec{x}\right)}=\frac{1}{2N\left(\vec{y}\right)}\frac{\delta S}{\delta B_{ij}\left(\vec{x}\right)\delta B_{ij}\left(\vec{y}\right)},
\end{eqnarray}
where we used (\ref{fd_pitld_Bij}) and (\ref{fd_pitld^ij_Bkl}).

From (\ref{PB_pitld_gen}),
\begin{eqnarray}
\left[\tilde{\pi}\left(\vec{x}\right),\tilde{\pi}\left(\vec{y}\right)\right] & = & -\frac{\delta\tilde{\pi}\left(\vec{y}\right)}{\delta N\left(\vec{x}\right)}+\frac{1}{N^{2}\left(\vec{x}\right)}\frac{\delta S}{\delta A\left(\vec{x}\right)}\frac{\delta\tilde{\pi}\left(\vec{y}\right)}{\delta\pi\left(\vec{x}\right)}-\frac{1}{N\left(\vec{x}\right)}\int\mathrm{d}^{3}z\frac{\delta^{2}S}{\delta A\left(\vec{x}\right)\delta N\left(\vec{z}\right)}\frac{\delta\tilde{\pi}\left(\vec{y}\right)}{\delta\pi\left(\vec{z}\right)}\nonumber \\
& = & \frac{1}{N\left(\vec{y}\right)}\frac{\delta^{2}S}{\delta N\left(\vec{x}\right)\delta A\left(\vec{y}\right)}-\frac{1}{N\left(\vec{x}\right)}\frac{\delta^{2}S}{\delta A\left(\vec{x}\right)\delta N\left(\vec{y}\right)},\label{PB_pi_tld_pi_tld}
\end{eqnarray}
where we used (\ref{fd_pitld_N}), and
\begin{eqnarray}
& & \left[\tilde{\pi}\left(\vec{x}\right),\tilde{\pi}^{ij}\left(\vec{y}\right)\right] \nonumber\\
& = & -\frac{\delta\tilde{\pi}^{ij}\left(\vec{y}\right)}{\delta N\left(\vec{x}\right)}-\frac{1}{N\left(\vec{x}\right)}\int\mathrm{d}^{3}z\frac{\delta^{2}S}{\delta A\left(\vec{x}\right)\delta h_{kl}\left(\vec{z}\right)}\frac{\delta\tilde{\pi}^{ij}\left(\vec{y}\right)}{\delta\pi^{kl}\left(\vec{z}\right)}\nonumber \\
& = & -\frac{1}{2}\delta^{3}\left(\vec{x}-\vec{y}\right)\frac{1}{N^{2}\left(\vec{y}\right)}\frac{\delta S}{\delta B_{ij}\left(\vec{y}\right)}+\frac{1}{2}\frac{1}{N\left(\vec{y}\right)}\frac{\delta^{2}S}{\delta N\left(\vec{x}\right)\delta B_{ij}\left(\vec{y}\right)}-\frac{1}{N\left(\vec{x}\right)}\frac{\delta^{2}S}{\delta A\left(\vec{x}\right)\delta h_{ij}\left(\vec{y}\right)},\qquad 
\end{eqnarray}
where we used (\ref{fd_pitld^ij_N}).

From (\ref{PB_pitld^ij_gen}), we get
\begin{eqnarray}
\left[\tilde{\pi}^{ij}\left(\vec{x}\right),\tilde{\pi}^{kl}\left(\vec{y}\right)\right] & = & -\frac{\delta\tilde{\pi}^{kl}\left(\vec{y}\right)}{\delta h_{ij}\left(\vec{x}\right)}-\frac{1}{2}\frac{1}{N\left(\vec{x}\right)}\int\mathrm{d}^{3}z\frac{\delta^{2}S}{\delta B_{ij}\left(\vec{x}\right)\delta h_{mn}\left(\vec{z}\right)}\frac{\delta\tilde{\pi}^{kl}\left(\vec{y}\right)}{\delta\pi^{mn}\left(\vec{z}\right)}\nonumber \\
& = & \frac{1}{2N\left(\vec{y}\right)}\frac{\delta^{2}S}{\delta h_{ij}\left(\vec{x}\right)\delta B_{kl}\left(\vec{y}\right)}-\frac{1}{2N\left(\vec{x}\right)}\frac{\delta^{2}S}{\delta B_{ij}\left(\vec{x}\right)\delta h_{kl}\left(\vec{y}\right)},
\end{eqnarray}
where we used (\ref{fd_pitld^ij_hkl}).

Finally, let us evaluate the Poisson brackets of the primary constraints with the canonical Hamiltonian $H_{\mathrm{C}}$.
Firstly, note since $p$, $p^{ij}$, $\tilde{\pi}$ and $\tilde{\pi}^{ij}$
have vanishing Poisson bracket with $N^{i}$, it immediately follows from (\ref{PB_LD_com}) that
	\begin{eqnarray}
	\left[p,X[\vec{N}]\right] & = & 0,\\
	\left[p^{ij},X[\vec{N}]\right] & = & 0,\\
	\left[\tilde{\pi},X[\vec{N}]\right] & = & 0,\\
	\left[\tilde{\pi}^{ij},X[\vec{N}]\right] & = & 0.
	\end{eqnarray}
The only non-vanishing one is $\left[\pi_{i}, X[\vec{N}]\right]$. Using (\ref{PB_LD_com})
we have, for a smoothing vector field $f^{i}\left(\vec{x}\right)$
(nothing to do with the phase space variables)
	\begin{eqnarray}
	\left[\int\mathrm{d}^{3}x\,f^{i}\left(\vec{x}\right)\pi_{i}\left(\vec{x}\right),X[\vec{N}]\right] & = & -X\left[\left[\vec{N},\int\mathrm{d}^{3}x\,f^{i}\left(\vec{x}\right)\pi_{i}\left(\vec{x}\right)\right]\right]\nonumber \\
	& = & -X[\vec{f}]\nonumber \\
	& \simeq & -\int\mathrm{d}^{3}x\,f^{i}\mathcal{C}_{i},\label{PB_f^i_pi_i_X}
	\end{eqnarray}
where in the last step we have used (\ref{X_xi_ibp}) by replacing $\vec{\xi} \rightarrow \vec{f}$. 
It immediately follows from (\ref{PB_f^i_pi_i_X}) that
	\begin{equation}
	\left[\pi_{i}\left(\vec{x}\right),X[\vec{N}]\right]\approx -\mathcal{C}_{i}\left(\vec{x}\right),\label{PB_pi_i_XNi}
	\end{equation} 
where $\mathcal{C}_{i}$ is defined in (\ref{calCi_def}).

It immediately follows from (\ref{PB_pi_i_XNi}) that\footnote{Instead of using (\ref{PB_LD_com}) and (\ref{PB_pi_i_XNi}), one can get the same result simply by definition
	\[
	\left[\pi_{i},H_{\mathrm{C}}\right]=-\frac{\delta H_{\mathrm{C}}}{\delta N^{i}}=-\frac{\delta}{\delta N^{i}}\left(\int\mathrm{d}^{3}x\left(NC\right)+X[\vec{N}]\right)\simeq-\frac{\delta}{\delta N^{i}}\int\mathrm{d}^{3}x\,N^{i}\mathcal{C}_{i}=-\mathcal{C}_{i}.
	\]}
	\begin{equation}
	\left[\pi_{i},H_{\mathrm{C}}\right]=\left[\pi_{i},\int\mathrm{d}^{3}x\left(NC\right)+X[\vec{N}]\right]=\left[\pi_{i},X[\vec{N}]\right]\approx-\mathcal{C}_{i}.
	\end{equation}
From (\ref{PB_p_gen}) and (\ref{Ham_C_fin}),
	\begin{equation}
	\left[p,H_{\mathrm{C}}\right]=\left[p,\int\!\mathrm{d}^{3}x\,NC\right]=-\frac{\delta}{\delta A}\int\!\mathrm{d}^{3}x\,NC = -N\tilde{\pi}\approx 0.
	\end{equation}
From (\ref{PB_p^ij_gen}) and (\ref{Ham_C_fin}),
	\begin{equation}
	\left[p^{ij},H_{\mathrm{C}}\right]=\left[p^{ij},\int\!\mathrm{d}^{3}x\,NC\right]=-\frac{\delta}{\delta B_{ij}}\int\!\mathrm{d}^{3}x\,NC=-2N\tilde{\pi}^{ij}\approx 0.
	\end{equation}
We also have
	\begin{eqnarray}
	\left[\tilde{\pi}\left(\vec{x}\right),H_{\mathrm{C}}\right] & = & \left[\tilde{\pi}\left(\vec{x}\right),\int\mathrm{d}^{3}y\,NC\right], \\
	\left[\tilde{\pi}^{ij}\left(\vec{x}\right),H_{\mathrm{C}}\right] & = &\left[\tilde{\pi}^{ij}\left(\vec{x}\right),\int\mathrm{d}^{3}y\,NC\right],
	\end{eqnarray}
then using (\ref{PB_pitld_gen}) and (\ref{PB_pitld^ij_gen}), after some manipulations we get (\ref{PB_pitld_HC}) and (\ref{PB_pitld^ij_HC}).

\section{Proof of $\left[\mathcal{C}(\vec{x}),\bar{p}(\vec{y}) \right] \approx 0$} \label{sec:PB_calC_pbar}

In this appendix we present the details in proving that $\left[\mathcal{C}\left(\vec{x}\right),\bar{p}\left(\vec{y}\right)\right]$
is weakly vanishing. To this end, it is more convenient to
consider the Poisson bracket $\left[\mathcal{C}[V],\bar{p}[U]\right]$, with
	\begin{equation}
	\mathcal{C}[V]:=\int\mathrm{d}^{3}x\,\mathcal{C}(\vec{x})V(\vec{x}),\qquad\bar{p}[U]:=\int\mathrm{d}^{3}x\,\bar{p}(\vec{x})U(\vec{x}),
	\end{equation}
where $V(\vec{x})$ and $U(\vec{x})$ are arbitrary smoothing functions.
Note $\bar{p}[U]$ has an equivalent expression (\ref{pbar_U_fn}).
A similar expression for $\mathcal{C}[V]$ can be found if we keep all the constraints (and thus keep equalities instead of weak equalities) in deriving (\ref{nev2_cc}), which yields
	\begin{eqnarray}
	&  & \int\mathrm{d}^{3}x\,\mathcal{V}_{\alpha}^{(2)}\left(\vec{x}\right)\left[\varphi^{\alpha}\left(\vec{x}\right),H_{\mathrm{C}}\right]\nonumber \\
	& = & \int\mathrm{d}^{3}x\,\left[\mathcal{V}_{\alpha}^{(2)}\left(\vec{x}\right)\varphi^{\alpha}\left(\vec{x}\right),H_{\mathrm{C}}\right]-\int\mathrm{d}^{3}x\,\left[\mathcal{V}_{\alpha}^{(2)}\left(\vec{x}\right),H_{\mathrm{C}}\right]\varphi^{\alpha}\left(\vec{x}\right)\nonumber \\
	& = & \left[\int\mathrm{d}^{3}x\,\bar{\pi}\left(\vec{x}\right)V\left(\vec{x}\right),H_{\mathrm{C}}\right]-\int\mathrm{d}^{3}x\,\left[\mathcal{V}_{\alpha}^{(2)}\left(\vec{x}\right),H_{\mathrm{C}}\right]\varphi^{\alpha}\left(\vec{x}\right)\nonumber \\
	& = & \int\mathrm{d}^{3}x\left[\bar{\pi}\left(\vec{x}\right),H_{\mathrm{C}}\right]V\left(\vec{x}\right)+\int\mathrm{d}^{3}x\left[V\left(\vec{x}\right),H_{\mathrm{C}}\right]\bar{\pi}\left(\vec{x}\right)-\int\mathrm{d}^{3}x\,\left[\mathcal{V}_{\alpha}^{(2)}\left(\vec{x}\right),H_{\mathrm{C}}\right]\varphi^{\alpha}\left(\vec{x}\right)\nonumber \\
	& \equiv & \int\mathrm{d}^{3}x\,\mathcal{C}\left(\vec{x}\right)V\left(\vec{x}\right)+\int\mathrm{d}^{3}x\left[V\left(\vec{x}\right),H_{\mathrm{C}}\right]\bar{\pi}\left(\vec{x}\right)-\int\mathrm{d}^{3}x\,\left[\mathcal{V}_{\alpha}^{(2)}\left(\vec{x}\right),H_{\mathrm{C}}\right]\varphi^{\alpha}\left(\vec{x}\right),
	\end{eqnarray}
together with
	\begin{equation}
	\int\mathrm{d}^{3}x\,\mathcal{V}_{\alpha}^{(2)}\left(\vec{x}\right)\left[\varphi^{\alpha}\left(\vec{x}\right),H_{\mathrm{C}}\right]\equiv\int\mathrm{d}^{3}x\,V_{a}\left(\vec{x}\right)\left[\tilde{\pi}^{a}\left(\vec{x}\right),H_{\mathrm{C}}\right],
	\end{equation}
we get
	\begin{eqnarray}
	\mathcal{C}[V] & := & \int\mathrm{d}^{3}x\,V_{a}(\vec{x})\left[\tilde{\pi}^{a}(\vec{x}),H_{\mathrm{C}}\right]\nonumber \\
	&  & -\int\mathrm{d}^{3}x\left[V(\vec{x}),H_{\mathrm{C}}\right]\bar{\pi}(\vec{x})+\int\mathrm{d}^{3}x\left[\mathcal{V}_{\alpha}^{(2)}(\vec{x}),H_{\mathrm{C}}\right]\varphi^{\alpha}(\vec{x}). \label{calC_V_fn}
	\end{eqnarray}
Although (\ref{pbar_U_fn}) and (\ref{calC_V_fn}) look more complicated, they are simpler in calculating the Poisson brackets.

Using the fact that $\left[\bar{\pi}\left(\vec{x}\right),\bar{p}[U]\right]\approx 0$ 
and $\left[\varphi^{\alpha}\left(\vec{x}\right),\bar{p}[U]\right]\approx 0$ (see (\ref{PB_p_Pbar})-(\ref{PB_pitld_Pbar})),
	\begin{eqnarray}
	&  & \left[\mathcal{C}[V],\bar{p}[U]\right]\nonumber \\
	& \equiv & \left[\int\mathrm{d}^{3}x\,V(\vec{x})\mathcal{C}(\vec{x}),\int\mathrm{d}^{3}\vec{y}\,U_{b}(\vec{y})p^{b}(\vec{y})\right]\nonumber \\
	& \approx & \left[\int\mathrm{d}^{3}x\,V_{a}(\vec{x})\left[\tilde{\pi}^{a}(\vec{x}),H_{\mathrm{C}}\right],\int\mathrm{d}^{3}\vec{y}\,U_{b}(\vec{y})p^{b}(\vec{y})\right]\nonumber \\
	& \approx & \int\mathrm{d}^{3}x\int\mathrm{d}^{3}y\left[V_{a}(\vec{x})\left[\tilde{\pi}^{a}(\vec{x}),H_{\mathrm{C}}\right],p^{b}(\vec{y})\right]U_{b}(\vec{y})\nonumber \\
	& = & \int\mathrm{d}^{3}x\int\mathrm{d}^{3}y\left\{ \left[V_{a}(\vec{x}),p^{b}(\vec{y})\right]\left[\tilde{\pi}^{a}(\vec{x}),H_{\mathrm{C}}\right]+V_{a}(\vec{x})\left[\left[\tilde{\pi}^{a}(\vec{x}),H_{\mathrm{C}}\right],p^{b}(\vec{y})\right]\right\} U_{b}(\vec{y}). \label{PB_VC_P_t1}
	\end{eqnarray}
Then using the Jacobi identity
	\begin{eqnarray}
	0 & \equiv & \left[\left[\tilde{\pi}^{a}\left(\vec{x}\right),H_{\mathrm{C}}\right],p^{b}\left(\vec{y}\right)\right]+\left[\left[H_{\mathrm{C}},p^{b}\left(\vec{y}\right)\right],\tilde{\pi}^{a}\left(\vec{x}\right)\right]+\left[\left[p^{b}\left(\vec{y}\right),\tilde{\pi}^{a}\left(\vec{x}\right)\right],H_{\mathrm{C}}\right]\nonumber \\
	& \approx & \left[\left[\tilde{\pi}^{a}\left(\vec{x}\right),H_{\mathrm{C}}\right],p^{b}\left(\vec{y}\right)\right]+\left[\left[p^{b}\left(\vec{y}\right),\tilde{\pi}^{a}\left(\vec{x}\right)\right],H_{\mathrm{C}}\right],
	\end{eqnarray}
we get
	\begin{equation}
	\left[\left[\tilde{\pi}^{a}\left(\vec{x}\right),H_{\mathrm{C}}\right],p^{b}\left(\vec{y}\right)\right]\approx\left[W^{ab}\left(\vec{x},\vec{y}\right),H_{\mathrm{C}}\right],
	\end{equation}
where we used the definition $W^{ab}\left(\vec{x},\vec{y}\right)\equiv\left[\tilde{\pi}^{a}\left(\vec{x}\right),p^{b}\left(\vec{y}\right)\right]$.
Plugging the above into (\ref{PB_VC_P_t1}) yields
	\begin{eqnarray}
	&  & \left[\mathcal{C}[V],\bar{p}[U]\right] \nonumber \\
	& \approx & \int\mathrm{d}^{3}x\int\mathrm{d}^{3}y\,\left\{ \left[V_{a}\left(\vec{x}\right),p^{b}\left(\vec{y}\right)\right]\left[\tilde{\pi}^{a}\left(\vec{x}\right),H_{\mathrm{C}}\right]+V_{a}\left(\vec{x}\right)\left[W^{ab}\left(\vec{x},\vec{y}\right),H_{\mathrm{C}}\right]\right\} U_{b}\left(\vec{y}\right)\nonumber \\
	& = & \int\mathrm{d}^{3}x\int\mathrm{d}^{3}y\,\big\{\left[V_{a}\left(\vec{x}\right),p^{b}\left(\vec{y}\right)\right]\left[\tilde{\pi}^{a}\left(\vec{x}\right),H_{\mathrm{C}}\right]\nonumber \\
	&  & \quad+\left[V_{a}\left(\vec{x}\right)W^{ab}\left(\vec{x},\vec{y}\right),H_{\mathrm{C}}\right]-\left[V_{a}\left(\vec{x}\right),H_{\mathrm{C}}\right]W^{ab}\left(\vec{x},\vec{y}\right)\big\} U_{b}\left(\vec{y}\right)\nonumber \\
	& \equiv & \int\mathrm{d}^{3}x\int\mathrm{d}^{3}y\,\left[\tilde{\pi}^{a}\left(\vec{x}\right),H_{\mathrm{C}}\right]\left[V_{a}\left(\vec{x}\right),p^{b}\left(\vec{y}\right)\right]U_{b}\left(\vec{y}\right)\nonumber \\
	& \equiv & \int\mathrm{d}^{3}x\,Z_{a}\left(\vec{x}\right)\left[\tilde{\pi}^{a}\left(\vec{x}\right),H_{\mathrm{C}}\right],\label{PB_VC_P_t2}
	\end{eqnarray}
where we used the fact that $V_{a}$ is the (left) null-eigenvector
of $W^{ab}$, and $U_{a}$ is the (right) null-eigenvector of $W^{ab}$.
In the last step of (\ref{PB_VC_P_t2}) we have defined
	\begin{equation}
	Z_{a}\left(\vec{x}\right):=\int\mathrm{d}^{3}\vec{y}\,\left[V_{a}\left(\vec{x}\right),p^{b}\left(\vec{y}\right)\right]U_{b}\left(\vec{y}\right),
	\end{equation}
for short.

Then there comes the crucial point. One can show that 
	\begin{equation}
	Z_{a}\left(\vec{x}\right)\propto V_{a}\left(\vec{x}\right).
	\end{equation}
To see this, we just need to check if
	\begin{equation}
	\int\mathrm{d}^{3}x\,Z_{a}\left(\vec{x}\right)W^{ac}\left(\vec{x},\vec{z}\right),
	\end{equation}
is vanishing or not. In fact,
	\begin{eqnarray}
	&  & \int\mathrm{d}^{3}x\,Z_{a}\left(\vec{x}\right)W^{ac}\left(\vec{x},\vec{z}\right)\nonumber \\
	& \equiv & \int\mathrm{d}^{3}x\,\left(\int\mathrm{d}^{3}y\,\left[V_{a}\left(\vec{x}\right),p^{b}\left(\vec{y}\right)\right]U_{b}\left(\vec{y}\right)\right)W^{ac}\left(\vec{x},\vec{z}\right)\nonumber \\
	& = & \int\mathrm{d}^{3}x\,\int\mathrm{d}^{3}y\,\left\{ \left[V_{a}\left(\vec{x}\right)W^{ac}\left(\vec{x},\vec{z}\right),p^{b}\left(\vec{y}\right)\right]-V_{a}\left(\vec{x}\right)\left[W^{ac}\left(\vec{x},\vec{z}\right),p^{b}\left(\vec{y}\right)\right]\right\} U_{b}\left(\vec{y}\right)\nonumber \\
	& = & -\int\mathrm{d}^{3}x\,\int\mathrm{d}^{3}y\,V_{a}\left(\vec{x}\right)\left[W^{ac}\left(\vec{x},\vec{z}\right),p^{b}\left(\vec{y}\right)\right]U_{b}\left(\vec{y}\right)\nonumber \\
	& \equiv & -\int\mathrm{d}^{3}x\,\int\mathrm{d}^{3}y\,V_{a}\left(\vec{x}\right)\left[\left[\tilde{\pi}^{a}\left(\vec{x}\right),p^{c}\left(\vec{z}\right)\right],p^{b}\left(\vec{y}\right)\right]U_{b}\left(\vec{y}\right),
	\end{eqnarray}
then using the Jacobi identity
	\begin{eqnarray}
	0 & \equiv & \left[\left[\tilde{\pi}^{a}\left(\vec{x}\right),p^{c}\left(\vec{z}\right)\right],p^{b}\left(\vec{y}\right)\right]+\left[\left[p^{c}\left(\vec{z}\right),p^{b}\left(\vec{y}\right)\right],\tilde{\pi}^{a}\left(\vec{x}\right)\right]+\left[\left[p^{b}\left(\vec{y}\right),\tilde{\pi}^{a}\left(\vec{x}\right)\right],p^{c}\left(\vec{z}\right)\right]\nonumber \\
	& = & \left[\left[\tilde{\pi}^{a}\left(\vec{x}\right),p^{c}\left(\vec{z}\right)\right],p^{b}\left(\vec{y}\right)\right]+\left[\left[p^{b}\left(\vec{y}\right),\tilde{\pi}^{a}\left(\vec{x}\right)\right],p^{c}\left(\vec{z}\right)\right],
	\end{eqnarray}
we get
	\begin{eqnarray}
	&  & \int\mathrm{d}^{3}x\,Z_{a}\left(\vec{x}\right)W^{ac}\left(\vec{x},\vec{z}\right)\nonumber \\
	& = & \int\mathrm{d}^{3}x\,\int\mathrm{d}^{3}y\,V_{a}\left(\vec{x}\right)\left[\left[p^{b}\left(\vec{y}\right),\tilde{\pi}^{a}\left(\vec{x}\right)\right],p^{c}\left(\vec{z}\right)\right]U_{b}\left(\vec{y}\right)\nonumber \\
	& = & -\int\mathrm{d}^{3}x\,\int\mathrm{d}^{3}y\,V_{a}\left(\vec{x}\right)\left[W^{ab}\left(\vec{x},\vec{y}\right),p^{c}\left(\vec{z}\right)\right]U_{b}\left(\vec{y}\right)\nonumber \\
	& = & -\int\mathrm{d}^{3}x\,\int\mathrm{d}^{3}y\,\left\{ \left[V_{a}\left(\vec{x}\right)W^{ab}\left(\vec{x},\vec{y}\right),p^{c}\left(\vec{z}\right)\right]-\left[V_{a}\left(\vec{x}\right),p^{c}\left(\vec{z}\right)\right]W^{ab}\left(\vec{x},\vec{y}\right)\right\} U_{b}\left(\vec{y}\right)\nonumber \\
	& = & 0,
	\end{eqnarray}
where again we used the fact that $V_{a}$ is the (left) null-eigenvector
of $W^{ab}$, and $U_{b}$ is the (right) null-eigenvector of $W^{ab}$.
The above result implies that $Z_{a}\left(\vec{x}\right)$ is also
a (left) null-eigenvector of $W^{ac}\left(\vec{x},\vec{z}\right)$.
On the other hand, since we assume the nullity of $W^{ab}$ is 1,
it has and only has one (left) null-eigenvector. We thus arrive at
the conclusion that
	\begin{equation}
	Z_{a}\left(\vec{x}\right)=c\,V_{a}\left(\vec{x}\right),\label{Z_cV}
	\end{equation}
with $c$ some numerical constant.

Finally, plugging (\ref{Z_cV}) into (\ref{PB_VC_P_t2}), we get
	\begin{equation}
	\left[\mathcal{C}[V],\bar{p}[U]\right]  \approx  \int\mathrm{d}^{3}x\,c\,V_{a}\left(\vec{x}\right)\left[\tilde{\pi}^{a}\left(\vec{x}\right),H_{\mathrm{C}}\right] 
	 \equiv  c\,\mathcal{C}[V]
	 \approx  0,
	\end{equation}
which implies
	\begin{equation}
	\left[\mathcal{C}\left(\vec{x}\right),\bar{p}\left(\vec{y}\right)\right] \approx 0.
	\end{equation}


\providecommand{\href}[2]{#2}\begingroup\raggedright\endgroup

\end{document}